\documentclass[
  journal=largetwo,
  manuscript=article-type,
  year=2020,
  volume=37,
]{cup-journal}

\usepackage{amsmath}
\usepackage{amssymb}
\usepackage[nopatch]{microtype}
\usepackage{booktabs}
\newcommand{\bl}[1]{\mbox{\boldmath$ #1 $}}

\title{C/O ratios in self-gravitating protoplanetary discs with dust evolution}

\author{Tamara Molyarova}
\affiliation{Institute of Astronomy, Russian Academy of Sciences, 48 Pyatnitskaya St., Moscow, 119017, Russia}
\alsoaffiliation{Research Institute of Physics, Southern Federal University, Stachki Ave. 194, Rostov-on-Don 344090, Russia}

\email[Tamara Molyarova]{moliarova@sfedu.ru}

\author{Eduard Vorobyov}
\affiliation{Institute of Astronomy, Russian Academy of Sciences, 48 Pyatnitskaya St., Moscow, 119017, Russia}
\alsoaffiliation{Research Institute of Physics, Southern Federal University, Stachki Ave. 194, Rostov-on-Don 344090, Russia}

\author{Vitaly Akimkin}
\affiliation{Institute of Astronomy, Russian Academy of Sciences, 48 Pyatnitskaya St., Moscow, 119017, Russia}


\begin{document}

\begin{abstract}
Elemental abundances, particularly the C/O ratio, are seen as a way to connect the composition of planetary atmospheres with planet formation scenario and the disc chemical environment.
We model the chemical composition of gas and ices in a self-gravitating disc on timescales of 0.5\,Myr since its formation to study the evolution of C/O ratio due to dust dynamics and growth, and phase transitions of the volatile species.
We use the thin-disc hydrodynamic code FEOSAD, which includes disc self-gravity, thermal balance, dust evolution and turbulent diffusion, and treats dust as a dynamically different and evolving component interacting with the gas. It also describes freeze-out, sublimation and advection of four most abundant volatile species: H$_2$O, CO$_2$, CH$_4$ and CO.
We demonstrate the effect of gas and dust substructures such as spirals and rings on the distribution of volatiles and C/O ratios, including the formation of multiple snowlines of one species, and point out the anticorrelation between dust-to-gas ratio and total C/O ratio emerging due to the contribution of oxygen-rich ice mantles. We identify time and spatial locations where two distinct trigger mechanisms for planet formation are operating and differentiate them by C/O ratio range: wide range of the C/O ratios of $0-1.4$ for streaming instability, and a much narrower range $0.3-0.6$ for gravitational instability (with the initial value of 0.34). This conclusion is corroborated by observations, showing that transiting exoplanets, which possibly experienced migration through a variety of disc conditions, have significantly larger spread of C/O in comparison with directly imaged exoplanets likely formed in gravitationally unstable outer disk regions. We show that the ice-phase $\mathrm{C/O}\approx0.2-0.3$ between the CO, CO$_2$ and CH$_4$ snowlines corresponds to the composition of the Solar system comets, that represent primordial planetesimals.
\end{abstract}

\section{Introduction}

The protoplanetary disc matter can be roughly divided into three component: gaseous chemical species, solid dust particles, and icy mantles covering the surface of dust grains. Gas and solid particles become dynamically decoupled, as evolving dust grows and acquires relative velocities leading to the redistribution of elements in the disc and between the phases, and creating the premises for different chemical environments. When planets start to form, their properties, including chemical composition of the atmosphere, are inevitably affected by the location and the mechanism of their formation. This suggests that the origin of (exo)planets might be investigated using their observed chemical composition, and makes understanding the disc chemical evolution vital for creating a consistent planet formation theory.

One of the key parameters that govern the chemical setup of a planetary atmosphere is the relation between the abundances of carbon and oxygen, often referred to as carbon-to-oxygen ratio (hereafter C/O ratio). The variations of C/O ratio in the ice and gas phases at the snowlines of main disc volatiles (CO, CO$_2$, and H$_2$O) and the prospects of connecting them to planet formation were discussed in \citet{2011ApJ...743L..16O} within a qualitative freeze-out model. Since then, C/O ratio received a lot of attention in this context. It was thoroughly investigated in modelling \citep[see, e.g.,][]{2017MNRAS.469.3994B,2018A&A...613A..14E,2019A&A...627A.127C,2019A&A...632A..63C,2020A&A...635A..68C,2020A&A...642A.229C,2020ApJ...899..134K,2021ApJ...909...40T,2021A&A...654A..71S}. The connection of disc chemical composition with C/O in exoplanetary atmospheres was modelled using core accretion model \citep{2015A&A...574A.138T} and ``chain'' planet population synthesis model \citep{2016ApJ...832...41M}. \citet{2022ApJ...934...74M} considered a simple formation retrieval pipeline and found that this task requires careful consideration of the model assumptions.

The measurements of molecular abundances in the atmospheres of giant exoplanets obtained by a variety of modern facilities, such as HST, Spitzer, VLTI, JWST, Gemini, indicate a diversity of C/O ratios: from low C/O ratio values~\citep[$\approx0.4$, below the solar value of $0.54$;][]{2019NatAs.tmp..361B,2020A&A...633A.110G,2024ApJ...964..168W,2024ApJ...963L...5X} to stellar \citep[$\approx0.5$, close to solar;][]{2020A&A...640A.131M,2021Natur.595..370Z,2024AJ....167..110S} and close to or above unity \citep{2009ApJ...704.1616S,2011Natur.469...64M}. A variety of solar and super-solar C/O ratios is observed in four planets within the HR8799 system \citep{2024arXiv240403776N}. 
Chemical composition of the atmospheres of many hot Jupiters indicates high C/O>1 of the forming material \citep{2013ApJ...763...25M}. There is an observational evidence of young planets in PDS~70 disc accreting the material with C/O>1 \citep{2021AJ....162...99F}. 
The number of exoplanets with constrained atmospheric C/O ratios grows with large studies of multiple planets such as \citet{2022ApJS..260....3C}, which allows us to make some statistical conclusions. The population study of C/O ratios in exoplanetary atmospheres reveals that there are two populations with different elemental ratio, which are likely formed in different mechanisms \citep{2023AJ....166...85H}. \citet{2023A&A...675A..95K} were able to restrict the formation scenario for WASP-77b based on the measured C/O ratio of the planet \citep{2021Natur.598..580L} and the modelling of planet formation and migration.

Elemental abundances in protoplanetary discs can be constrained from observations \citep{2020A&A...638A.110F}, and C/O ratio in the gas can be estimated. Spatially resolved observations can help distinguish between different C/O ratios spectroscopically \citep{2020MNRAS.497.2540M}. \citet{2018ApJ...865..155C} report ${\rm C/O}\approx0.8$ in the molecular layer of IM~Lup disc. ALMA observations of hydrocarbons and sulphur-bearing species indicate $\mathrm{C/O}>1$ in the upper disc layers and in the outer disc in TW~Hya and DM~Tau \citep{2011A&A...535A.104D,2016ApJ...831..101B,2018A&A...617A..28S} and for a population of discs in Lupus \citep{2019A&A...631A..69M}. For the nearby discs the solar elemental composition with $\mathrm{C/O}\approx0.54$ is usually expected, thus the observed higher values confirm redistribution of carbon and oxygen in discs. In addition to high C/O in disc atmospheres, there is evidence of both carbon and oxygen depletion from gas \citep{2016A&A...592A..83K,2019A&A...631A..69M}. However, some of heavy oxygen carriers might not be observable, leading to overestimated C/O in disc observations \citep{2015A&A...582A..88W}.

The volatile composition is also used to constrain the origin of bodies in the Solar System. Fraction of CO and CO$_2$ ices relative to water in cometary comae indicate their formation between the CO and CO$_2$ snowlines or exterior to the CO snowline \citep{2012ApJ...758...29A,2022PSJ.....3..150S}. Abundances of CO and N$_2$ ices were used to analyse the original location of Pluto and Triton \citep{2024arXiv240603815M}.
Observed elemental abundances were used to constrain the Jupiter formation scenario \citep{2004ApJ...611..587L}, relying also on abundances of nitrogen~\citep{2019AJ....158..194O,2019A&A...632L..11B} and chemically inactive species like~Ar. However, the model assumptions can lead to different interpretation of the observations: while \citet{2019AJ....158..194O} and \citet{2019A&A...632L..11B} suggest that Jupiter formed outside N$_2$ snowline (at $>30$\,au), \citet{2021A&A...651L...2O} consider the concept disc shadow, which allows Jupiter to form near its current location.

Chemical processes other than freeze-out and sublimation at the snowlines can alter the composition of ice and gas as well. Due to gas-phase and surface reactions, snowlines can become important for the redistribution of elements. More detailed chemical modelling shows that the C/O ratio in  the gas and in the ice depends also on the initial chemical setup and ionisation by cosmic rays and radioactive nuclei \citep{2016A&A...595A..83E,2018A&A...613A..14E}. It directly affects the interpretation of observations. Another essential chemical process is CO depletion from the gas, resulting in its transformation to CO$_2$ ice \citep{2018A&A...611A..80B}. For example, stellar C/O ratio in the atmosphere of HR\,8799e indicates that the planet accreted its material beyond CO snowline ($\approx$45\,au), but chemical modelling suggests that due to CO depletion, the C/O in the ice already approaches the stellar ratio beyond CO$_2$ snowline ($\approx$20\,au), which is closer to the star \citep{2020A&A...640A.131M}.

Another key process affecting the elemental ratio is dust drift, which leads to spatial segregation between the chemical constituents of the gas and the grains covered with ice. The distribution of CO in the gas and ice phases was studied within dynamical models of dust evolution \citep{2017A&A...600A.140S,2018ApJ...864...78K}. Even without chemical processes, dust evolution and dynamics can substantially alter C/O ratio in the atmospheres of forming planets \citep{2017MNRAS.469.3994B}. Some models combine chemical reactions treatment with dust evolution and transport, usually within 1D viscous models. Dust transport can have a strong effect on the abundances of volatiles in the inner disc regions \citep{2018A&A...611A..80B}. However, for discs with low turbulence and high cosmic ray ionisation rate, C/O ratio is rather defined by chemical evolution \citep{2019MNRAS.487.3998B}.

In our previous work \citep{2021ApJ...910..153M} we showed that the volatile species tend to concentrate around their snowlines both in the gas and more notably on the dust surface. This accumulation was found to be caused by effective transport of volatiles through the snowlines by azimuthal variations in the gas and dust radial and angular velocity, an effect that cannot be captured in 1D viscous disc models. Such accumulation should immediately affect the local C/O ratio, which suggests the connection between the snowlines of various volatiles and the formation of planets with altered C/O in their atmospheres. In this work, we follow the distribution of the main volatile species in the disc to investigate the distribution of C/O ratio in gas and ice in a 2D thin-disc hydrodynamic model. We study the effect of dust growth and dynamics on the elemental ratios and consider the role of the initial mass of the collapsing core on the distribution of volatiles.

The paper is organised as follows. The main features of the used FEOSAD model are described in Section~\ref{sec:model}, with the details of the treatment of the volatiles given in Section~\ref{sec:chemistry}. In Section~\ref{sec:results} we describe the results of the simulations, focusing on distribution of the volatiles in Section~\ref{sec:volatiles_dist}, the C/O ratios in Section~\ref{sec:C2O_ratio}, and their evolution in Section~\ref{sec:snowline_evolution}. In Section~\ref{sec:discussion}, we discuss the implications of our results in the context of planet formation via different mechanisms. The main conclusions are listed in Section~\ref{sec:conclusions}.

\section{Model}
\label{sec:model}

We use the global model of protoplanetary disc formation FEOSAD~\citep{2018A&A...614A..98V}, which includes disc self-gravity, dust evolution and interaction with gas (including backreaction of dust on gas), turbulent viscosity, adiabatic and radiative cooling and heating. It describes the formation of a protostar and a protoplanetary disc from a collapsing cloud in a 2D thin-disc approach. The model also includes freeze-out of main volatile species as in \citet{2021ApJ...910..153M}, with the feedback from ice mantles on dust evolution via fragmentation velocity. Here we summarise the key characteristics of the model, more details can be found in the previous works \citep{2018A&A...614A..98V,2021ApJ...910..153M,2022MNRAS.516.4448K}.
The main difference from our previous study in \citet{2021ApJ...910..153M} is that here we consider the formation of dead zones via variable $\alpha$-parameter of Shakura and Sunyaev and also include turbulent diffusion.

\subsection{Gas evolution}
\label{sec:gas_evolution}

For the gas component, the hydrodynamic equations for mass, momentum, and internal energy conservation are the following
\begin{equation}
\label{cont}
\frac{{\partial \Sigma_{\rm g} }}{{\partial t}} + \nabla \cdot 
\left( \Sigma_{\rm g} \bl{v} \right) =0, 
\end{equation}
\begin{eqnarray}
\label{mom}
\frac{\partial}{\partial t} \left( \Sigma_{\rm g} \bl{v} \right) + [\nabla \cdot \left( \Sigma_{\rm
g} \bl{v} \otimes \bl{v} \right)] & =& - \nabla {\cal P} + \Sigma_{\rm g} \, \bl{g} + \nonumber
\\ 
+ \nabla \cdot \bl{\Pi} - \Sigma_{\rm d,gr} \bl{f},
\end{eqnarray}
\begin{equation}
\frac{\partial e}{\partial t} +\nabla \cdot \left( e \bl{v} \right) = -{\cal P} 
(\nabla \cdot \bl{v}) -\Lambda +\Gamma + 
\nabla \bl{v}:\Pi,
\label{eq:energy}
\end{equation}
where subscripts $p$ and $p^\prime$ denote the planar components $(r,\phi)$ in polar coordinates, $\Sigma_{\rm g}$ is the gas mass surface density, $e$ is the internal energy per surface area, ${\cal P}$ is the vertically integrated gas pressure calculated via the ideal equation of state as ${\cal P}=(\gamma-1) e$ with $\gamma=7/5$, $f$ is the friction force between gas and dust, $\bl{v}=v_r \hat{\bl r}+ v_\phi \hat{\bl \phi}$ is the gas velocity in the disc plane, and  $\nabla=\hat{\bl r} \partial / \partial r + \hat{\bl \phi} r^{-1} \partial / \partial \phi $ is the gradient along the planar coordinates of the disc. The gravitational acceleration in the disc plane, $\bl{g}=g_r \hat{\bl r} +g_\phi \hat{\bl \phi}$,  includes the gravity of the central protostar when formed and takes into account disc self-gravity of both gas and dust found by solving the Poisson integral \citep{1987gady.book.....B}.

The consideration of time-dependent energy balance (Eq.~\eqref{eq:energy}) allows us to accurately calculate the midplane temperature $T_{\rm mp}$ and is particularly important to describe the phase state of the volatiles and the level of turbulent viscosity. The terms $\Lambda$ and $\Gamma$ describe the rates of dust cooling and heating, respectively, by stellar and background irradiation. They are calculated  based on the analytical solution of the radiation transfer equations in the vertical direction  \citep{2016ApJ...823..141D,2018A&A...614A..98V} 
\begin{equation}
\Lambda = \frac{8 \tau_{\rm P}\sigma T_{\rm mp}^4}{1+2\tau_{\rm P}+\frac{3}{2}\tau_{\rm P}\tau_{\rm R}}, \text{    } \Gamma = \frac{8 \tau_{\rm P}\sigma T_{\rm irr}^4}{1+2\tau_{\rm P}+\frac{3}{2}\tau_{\rm P}\tau_{\rm R}}.
\label{eq:gammalambda}
\end{equation}
Here, $\sigma$ is the Stefan-Boltzmann constant, $\tau_{\rm P}$ and $\tau_{\rm R}$ are the Planck and Rosseland mean optical depths to the disc midplane, calculated as $\tau=\kappa \Sigma_{\rm dust}$ from  Planck and Rosseland mean opacities $\kappa_{\rm P}$ and $\kappa_{\rm R}$ \citep{2003A&A...410..611S} and total dust surface density $\Sigma_{\rm dust}$.  Gas and dust temperatures are assumed to be equal, and the midplane temperature is linked with gas pressure as $T_{\rm mp} = {\cal P}\mu/{\cal R}\Sigma_{\rm g}$, where $\mu=2.3$ is the mean molecular weight of the gas, and ${\cal R}$ is the universal gas constant. The  irradiation temperature at the disc surface $T_{\rm irr}$ is determined by both stellar and background irradiation. Stellar irradiation includes the luminosity from the photosphere of the protostar and accretion luminosity. The background radiation is assumed as a black body with the temperature of 15~K. For more details on the irradiation we refer to \citet{2018A&A...614A..98V}.

Turbulent viscosity is described using the common $\alpha$-parameter approach of \citet{1973A&A....24..337S}. It is taken into account via the viscous stress tensor $\bl{\Pi}$ \citep[see][for explicit expressions for the components of the terms with $\bl{\Pi}$]{2010ApJ...719.1896V}. The magnitude of kinematic viscosity is $\nu=\alpha c_{\rm s} H$, where $c_{\rm s}$ is the sound speed and $H$ is the vertical scale height of the gas disc calculated using an assumption of local hydrostatic equilibrium of a self-gravitating disc~\citep[see][Appendix~A]{2010ApJ...719.1896V}. Here, we use the adaptive $\alpha$ approach implying accretion through a layered disc~\citep{1996ApJ...457..355G,2001MNRAS.324..705A,2022MNRAS.516.4448K}. Turbulence is assumed to be generated by magneto-rotational instability (MRI) which only develops in layers of the disc where the ionisation level is high enough. The MRI-active layer is characterised by its surface density $\Sigma_{\rm MRI}$ and relatively high value of turbulent viscosity $\alpha_{\rm MRI}=10^{-3}$. As thermal and photo-ionisation are not efficient enough for the relatively cold and dense matter in  the disc at >0.5\,au, the main process determining the thickness of the MRI-active layer is ionisation by cosmic rays. It is assumed to be constant $\Sigma_{\rm MRI}=100$\,g~cm$^{-2}$, which is the typical depth of Galactic cosmic rays penetration in the ISM~\citep{1981PASJ...33..617U} and in protoplanetary  discs~\citep{2009A&A...501..619P}. The dead zone is characterised by surface density from the midplane $\Sigma_{\rm dz}=\Sigma_{\rm g}/2-\Sigma_{\rm MRI}$ with residual turbulence $\alpha_{\rm dz}$. The turbulence in this layer is only hydrodynamic turbulence driven by
the Maxwell stress in the active layer, and small value $\alpha_{\rm dz}=10^{-5}$ is adopted~\citep{2011ApJ...742...65O}. However, if local temperature exceeds the critical value of 1300\,K, thermal ionisation becomes possible, the MRI develops and the dead zone is no longer dead, in which case $\alpha_{\rm dz}=10^{-1}$ \citep{2010ApJ...713.1134Z,2019ApJ...882...96K}. This value is higher than $\alpha_{\rm MRI}$ in the outer disc due to the different ionisation processes and local conditions. In the outer disc, the MRI can be suppressed by non-ideal magneto-hydrodynamic (MHD) effects such as ambipolar diffusion and Ohmic resistivity \citep{2013ApJ...769...76B,2015ApJ...801...84G}. In the dead zone in the inner disc, $\alpha$ can reach higher values when the MRI is triggered by thermal ionisation, as shown by 3D MHD simulations \citep{2020MNRAS.495.3494Z}.
The adopted parameterization makes use of an effective parameter $\alpha_{\rm eff}$, which at any given location is calculated as
\begin{equation}
    \alpha_{\rm eff}=\frac{\Sigma_{\rm MRI}\alpha_{\rm MRI}+\Sigma_{\rm dz}\alpha_{\rm dz}}{\Sigma_{\rm MRI}+\Sigma_{\rm dz}}, \\
    \alpha_{\rm dz}=
\begin{cases}
10^{-5}, & \text{ if } T_{\rm mp} < 1300 \text{K}; \\
10^{-1}, & \text{ if } T_{\rm mp} \geq 1300 \text{K}.
\end{cases}
\label{eq:alpha}
\end{equation}

\subsection{Dust evolution}
\label{sec:dust_evolution}

Dust is described as consisting of two components: small grains that are dynamically coupled to the gas, with the mass surface density $\Sigma_{\rm d,sm}$, and grown grains with the mass surface density $\Sigma_{\rm d,gr}$ that can move relative to the gas and change in size. The total dust surface density necessary for the calculation of the optical depths for Eq.~\eqref{eq:gammalambda} is  $\Sigma_{\rm dust}=(\Sigma_{\rm d,gr}+\Sigma_{\rm d,sm})/2$. The factor $1/2$ appears as the optical depths is calculated to the midplane. Each dust population has a power-law size distribution  $f(a)= dN/da = C a^{-p}$ with a normalisation constant $C$ and a fixed exponent $p=3.5$. Small dust has sizes between $a_{\rm min}=5\times 10^{-7}$\,cm and  $a_*=10^{-4}$\,cm, grown dust has sizes between $a_*$ and $a_{\rm max}$, which can vary due to dust coagulation and fragmentation. Dynamics of these dust components follows the continuity and momentum equations 
\begin{equation}
\label{contDsmall}
\frac{{\partial \Sigma_{\rm d,sm} }}{{\partial t}}  + \nabla  \cdot 
\left( \Sigma_{\rm d,sm} \bl{v} \right) = - S(a_{\rm max}) + \nabla  \cdot \left( D \Sigma_{\rm g} \nabla\left(\frac{\Sigma_{\rm d,sm}}{\Sigma_{\rm g}} \right)\right),  
\end{equation}
\begin{equation}
\label{contDlarge}
\frac{{\partial \Sigma_{\rm d,gr} }}{{\partial t}}  + \nabla  \cdot 
\left( \Sigma_{\rm d,gr} \bl{u} \right) = S(a_{\rm max}) + \nabla  \cdot \left( D \Sigma_{\rm g} \nabla\left(\frac{\Sigma_{\rm d,gr}}{\Sigma_{\rm g}} \right)\right),  
\end{equation}
\begin{eqnarray}
\label{momDlarge}
\frac{\partial}{\partial t} \left( \Sigma_{\rm d,gr} \bl{u}\right) +  \left[\nabla \cdot \left( \Sigma_{\rm
d,gr} \bl{u} \otimes \bl{u} \right)\right]  &=&   \Sigma_{\rm d,gr} \, \bl{g} + \nonumber \\
 + \Sigma_{\rm d,gr} \bl{f} + S(a_{\rm max}) \bl{v},
\end{eqnarray}
where $\bl{u}$ is the grown dust velocity. The term $S(a_{\rm max})$ is responsible for the exchange of matter between the dust populations, as dust is converted from the small to grown component due to coagulation and back due to fragmentation. The details of the dust evolution model are presented in~\citet{2018A&A...614A..98V}. The last term in Eqs.~\eqref{contDsmall} and~\eqref{contDlarge} is responsible for dust turbulent diffusion, similar to \citet{2020A&A...643A..13V}. The coefficient of turbulent diffusion $D$ is related to the kinematic viscosity $\nu$ as $D=\nu/(1+\mathrm{St}^2)$ \citep{2023arXiv231213287B}. Diffusion affects dust grains along with their ice mantles, as well as the gas-phase species (see Section~\ref{sec:chemistry}).

The innermost regions of the disc are challenging to simulate explicitly due to the Courant criterion: in the highly dynamic inner regions (fraction of au) the timescales are so short that the code demands very small time step in order to preserve stability. Therefore, the inner regions are represented by a sink cell, with a carefully chosen inflow-outflow boundary condition at the sink cell and a parametric description of the accretion onto the star~\citep[see][for details]{2018A&A...614A..98V}. In the simulations presented below, the radius of the sink cell is 0.62\,au.

We consider two disc models with different initial mass of the collapsing cloud, 0.66 and 1~$M_{\odot}$. We note that about 10\% of the gas mass that crosses the sink cell is assumed to be evacuated by jets and outflows, and the other 90\% lands on to the star. A small amount of mass remains in the envelope by the end of simulations.  In both models, the initial gas temperature $T_{\rm init}=15$\,K  and the ratio of rotational to gravitational energy $\beta=0.28$\%. The simulations start from the collapse of a molecular cloud, with only small dust grains. The simulation continues until the age of the system becomes equal to 0.5\,Myr. Masses of the central protostar and the disc by the end of simulation are $M_{\star} = 0.4$\,$M_{\odot}$ and  $M_{\rm disc} = 0.22$\,$M_{\odot}$ in model~M1 and $M_{\star} = 0.58$\,$M_{\odot}$ and $M_{\rm disc} = 0.35$\,$M_{\odot}$ in model~M2. The disc masses are around $0.5$ stellar masses, which makes them essentially self-gravitating.

A number of recent studies develop the idea that the mass infall from the ambient ISM continues during the lifetime of the disc, including the Class~II stage \citep{2024arXiv240507334P,2024arXiv240506520P,2024arXiv240508451W}. 
Such models describe a Bondi--Hoyle accretion regime and are in good agreement with the observed properties of the disc population, such as accretion rates, masses and sizes. This input of matter can play an important role in disc evolution and  planet formation process \citep{2015A&A...573A...5V,2016A&A...587A.146V}. In the FEOSAD model, the mass infall to the disc can be accounted for \citep{2015A&A...573A...5V,2016A&A...587A.146V}, but in the present simulation this effect is not included. The mass infall from the envelope continues until the cloud is depleted of matter. Because of the thin-disc geometry, the gravitational contraction of the cloud proceeds in the plane of the disc. The matter is landing on the disc outer edge and is transported towards the star by the combined action of gravitational and viscous torques. The infall on the disc inner regions is therefore neglected. This is a reasonable approximation, considering that most of the matter and angular momentum in a three-dimensional cloud is located at relatively large polar angles and a flared outer edge of the disc intercepts most of them \citep[see, e.g.,][Figure~1]{2009A&A...495..881V}. In our modelling, we do not consider the continuous Bondi-Hoyle accretion and concentrate on the internal disc processes.

\subsection{Evolution of volatiles}
\label{sec:chemistry}

We follow the evolution of four main volatile species: H$_2$O, CO$_2$, CO and CH$_4$. These are the most abundant carbon- and oxygen-bearing ices observed in protostellar cores~\citep{2011ApJ...740..109O}. In the model, each of these species can be present in three states: in the gas, in the ice on the surface of small dust, and in the ice on the surface of grown dust. Each species $s$ is described by its surface density in the gas $\Sigma_{s}^{\rm gas}$, on small dust $\Sigma_{s}^{\rm sm}$, and on grown dust $\Sigma_{s}^{\rm gr}$. Their distributions in the disc  can change through three main processes: advection together with the corresponding component (gas or small/grown dust); exchange of mantles between dust populations due to grain collisions; phase transitions, including adsorption from gas to dust, and thermal and photo-desorption. Initially, all ices are on small grains and no volatiles present in the gas. The treatment of volatiles is adopted from~\citet{2021ApJ...910..153M}, who describe the models in more details. Here we recap main features of the chemical model.

Our chemical model only describes phase transitions, i.e. adsorption and desorption, which includes thermal desorption and photo-desorption by interstellar UV radiation. These reactions were shown to be he most important for gas-phase abundances of most species \citep{2011MNRAS.417.2950I}. Due to high computational costs, no other chemical processes, either gas-phase or surface reactions are included, although they also may have significant effect on the composition of both ice and gas~\citep{2011ApJS..196...25S}. The chemical evolution of the surface densities of volatile species is calculated from the system of equations
 \begin{eqnarray}
 \frac{{\partial \Sigma_{s}^{\rm gas} }}{{\partial t}}  + \nabla  \cdot 
 \left(\Sigma_{s}^{\rm gas} \bl{v} \right)  - \nabla  \cdot \left( D \Sigma_{\rm g} \nabla\left(\frac{\Sigma_{s}^{\rm gas} }{\Sigma_{\rm g}} \right)\right) &=&  \nonumber \\-\lambda_s\Sigma_{s}^{\rm gas}+\eta_s^{\rm sm}+\eta_s^{\rm gr},\label{eq:sig1}\\
 \frac{{\partial \Sigma_{s}^{\rm sm} }}{{\partial t}}  + \nabla  \cdot 
 \left(\Sigma_{s}^{\rm sm} \bl{v} \right) - \nabla  \cdot \left( D \Sigma_{\rm g} \nabla\left(\frac{\Sigma_{s}^{\rm sm}}{\Sigma_{\rm g}} \right)\right)&=& \nonumber \\ \lambda_s^{\rm sm}\Sigma_{s}^{\rm gas}-\eta_s^{\rm sm} ,\label{eq:sig2}\\
 \frac{{\partial \Sigma_{s}^{\rm gr} }}{{\partial t}}  + \nabla  \cdot 
 \left( \Sigma_{s}^{\rm gr} \bl{u} \right) -\nabla  \cdot \left( D \Sigma_{\rm g} \nabla\left(\frac{\Sigma_{s}^{\rm gr}}{\Sigma_{\rm g}} \right)\right) &=& \nonumber \\ \lambda_s^{\rm gr}\Sigma_{s}^{\rm gas}-\eta_s^{\rm gr},\label{eq:sig3}
 \end{eqnarray}
 where the mass rate coefficients per disc unit area of adsorption $\lambda_{s}$ and desorption $\eta_{s}$ for the species $s$ are calculated for local conditions at every hydrodynamic step, separately for small and grown dust populations. Eqs.~\eqref{eq:sig1}--\eqref{eq:sig3} are solved in two steps: first, the right-hand side is considered without the advection term. The case of pure adsorption/desorption represented by the right-hand side of the equation can be solved analytically~\citep[see Appendix~A in][]{2021ApJ...910..153M}. This is done at every hydrodynamic step, before the dust growth step, when ices on small and grown grains are redistributed proportionally to mass exchange between the dust populations. This is followed by a transport step, when the surface densities of the volatiles are changed according to the fluxes of their respective gas or dust components between the cells. Restricting chemical processes to only adsorption and desorption allows the chemical step to be calculated fast, which is very important for computationally demanding hydrodynamic simulations.

For each dust population, the desorption rate is a sum of thermal desorption and photo-desorption $\eta=\eta_{\rm td}+\eta_{\rm pd}$ (the indices ``sm'' and ``gr'' are omitted for convenience). We operate under the assumption of zeroth-order desorption, i.e. the desorption rate does not depend on the present amount of ice ($\Sigma_s^{\rm sm}$ or $\Sigma_s^{\rm gr}$). It implies that only the upper layers of the ice mantle are able to sublimate, which is a more appropriate approach for thick mantles. This assumption better describes desorption of CO and H$_2$O in temperature programmed desorption (TPD) experiments~\citep{2001MNRAS.327.1165F,2005ApJ...621L..33O,2006A&A...449.1297B} than first-order desorption, which is more suitable for thin mantles of several monolayers. The rate of thermal desorption is calculated as
\begin{equation}
\label{eq:desorption_rate}
\eta_{\rm td} =  \widetilde{\sigma}_{\rm tot} N_{\rm ss} \mu_{\rm s} m_{\rm p} \sqrt{\frac{2 N_{\rm ss} E_{\rm b}k_{\rm B}}{\pi^2 \mu_{\rm s} m_{\rm p}}} \exp{\left(-\frac{E_{\rm b}}{T_{\rm mp}}\right)},
\end{equation}
where $N_{\rm ss}=10^{15}$\,cm$^{-2}$ is the surface density of binding sites~\citep{2017SSRv..212....1C}, $E_{\rm b}$ (K) is the binding energy of the species to the surface, $\mu_{\rm s}$ is the species molecular mass, $m_{\rm p}$ is atomic mass unit, $k_{\rm B}$ is the Boltzmann constant. We follow \citet{1993MNRAS.263..589H} and use binding energy to calculate the pre-exponential factor in Eq.~\eqref{eq:desorption_rate}, the same way it was done in~\citet{2021ApJ...910..153M}.  However, this approach was recently demonstrated by~\citet{2022ESC.....6..597M} to underestimate the pre-exponential factor by a few orders of magnitude, as it does not account for the rotational part of the partition functions of desorbing molecules. Total surface area of dust grains (small or grown) per disc unit area $\widetilde{\sigma}_{\rm tot}$ (cm$^2$\,cm$^{-2}$) is calculated for the adopted power-law size distribution with $p=3.5$ as
\begin{eqnarray}
\label{eq:surface}
\widetilde{\sigma}_{\rm tot}^{\rm sm}&=&\frac{3 \Sigma_{\rm d, sm}}{\rho_{\rm s}\sqrt{a_{\rm min} a_*}},\\
\label{eq:surface1}
\widetilde{\sigma}_{\rm tot}^{\rm gr}&=&\frac{3 \Sigma_{\rm d, gr}}{\rho_{\rm s}\sqrt{a_* a_{\rm max}}}.
\end{eqnarray}
 Here, $\rho_{\rm s}=3$\,g~cm$^{-3}$ is the material density  of silicate cores of the dust grains. 

The model includes photodesorption of volatiles by interstellar irradiation, which is mostly relevant in the outer disc regions with lower optical depth. We do not consider the UV radiation field produced by the star and the accretion region as a source of photodesorption, assuming that they do not reach disc midplane due to high optical depth. The photo-desorption rate is calculated as
\begin{equation}
\label{eq:photodes}
     \eta_{\rm pd} =\widetilde{\sigma}_{\rm tot} \mu_{\rm sp}m_{\rm p}Y F_{\rm UV}.
\end{equation}
where $F_{\rm UV}=F_0^{\rm UV}G_{\rm UV}$ (photons~cm$^{-2}$\,s$^{-1}$) is the UV photon flux expressed in the units of standard UV field and $Y = 3.5 \times 10^{-3} + 0.13  \exp\left({-336\mathrm{K}/T_{\rm mp}}\right)$ (mol~photon$^{-1}$) is the photodesorption yield adopted from \citet{1995Natur.373..405W} for water ice. The intensity of the interstellar UV radiation field with $G_0=1$ is $F^{\rm UV}_0=4.63\times10^7$\,photon~cm$^{-2}$\,s$^{-1}$~\citep{1978ApJS...36..595D}. We assume that the disc situated in a star-forming region is illuminated by a slightly elevated unattenuated UV field with $G_{\rm env}=5.5G_0$. For the disc midplane, which is illuminated from above and below, this field scales with the UV optical depth $\tau_{\rm UV}$ towards the disc midplane as
 \begin{equation}
 G_{\rm UV}=0.5 G_{\rm env} e^{-\tau_{\rm UV}}.
 \label{eq:gfactoruv}
 \end{equation}
 The optical depth can be calculated as
 $\tau_{\rm UV}= 0.5 (\varkappa_{\rm sm} \Sigma_{\rm d,sm} + \varkappa_{\rm gr} \Sigma_{\rm d,gr})$, where $\varkappa_{\rm sm}=10^4$\,cm$^2$\,g$^{-1}$, $\varkappa_{\rm gr}=2\times 10^2$\,cm$^2$\,g$^{-1}$ are typical values of absorption coefficients in the UV for small and grown grains~\citep[Fig.~1]{2019MNRAS.486.3907P}.

We calculate the adsorption rate $\lambda$ following \citet{1990MNRAS.244..432B}. It is proportional to the total cross-section of dust grains per unit volume, which can be obtained from  the total surface area of dust grains per unit disc surface $\widetilde{\sigma}_{\rm tot}$. To change the normalisation to the 2D case, $\widetilde{\sigma}_{\rm tot}$ needs to be multiplied by $1/2H$. Another factor of $1/4$ follows from the difference between cross-section and surface area of a sphere. As a result, the adsorption  rate is calculated as
\begin{equation}
\lambda= \frac{\widetilde{\sigma}_{\rm tot}}{8H} \sqrt{\frac{8k_{\rm B}T_{\rm mp}}{\pi \mu_{\rm sp} m_{\rm p}}}.
 \label{eq:lambda}
 \end{equation}
 A more detailed derivation of rate coefficients for  adsorption and desorption is presented in Section~2.3 of~\citet{2021ApJ...910..153M}.

\begin{table}
 \caption{Binding energies, molecular weights, and initial abundances for the considered volatiles adopted in the modelling. Initial abundances of the species $f_{\rm s}$ are shown relative to number density of water molecules in ice phase, and $\Sigma_{s}^{\rm sm}/\Sigma_{\rm g}^{\rm init}$ is the corresponding initial mass fraction of the ices relative to gas surface density.}
 \label{tab:abundances}
 \centering
  \begin{tabular}{lcccc}
  \hline\noalign{\smallskip}
 Species  & $E_{\rm b}$, K & $\mu_{\rm s}$, amu  & $f_{\rm s}$, \% & $\Sigma_{s}^{\rm sm}/\Sigma_{\rm g}^{\rm init}$ \\
     \hline\noalign{\smallskip}
 H$_2$O & 5770 & 18 & 100 & $3.90\times10^{-4}$  \\
 CO$_2$ & 2360 & 44 & 29  & $2.77\times10^{-4}$  \\
 CO     & 850  & 28 & 29  & $1.76\times10^{-4}$  \\
 CH$_4$ & 1100 & 16 & 5   & $1.74\times10^{-5}$  \\
   \noalign{\smallskip}\hline
 \end{tabular}
 \end{table}

Table~\ref{tab:abundances} summarises the molecular parameters used in both models in this work. Binding energies for H$_2$O, CO$_2$, and CO are based on the experimental data from \citet{2017SSRv..212....1C} for desorption from crystalline water ice. The binding energy for methane is taken from \citet{1996ApJ...467..684A}. The values of the initial abundances relative to water $f_{\rm s}$ are based on the observations of ices in protostellar cores around the low-mass protostars \citep{2011ApJ...740..109O}. They are transformed to the mass fraction relative to gas assuming the water abundance of $5\times10^{-5}$ relative to gas number density. Total initial mass of ices comprises $\approx 8.5\%$ of the total mass of refractory grain cores. This is relatively low compared to the estimates suggesting comparable masses of ices and refractories in the discs \citep{2014prpl.conf..363P}. However, the lower fraction of ices is more suitable in our approach, that suggests  that ice mantles do not change mass and radius of dust grains \citep{2021ApJ...910..153M}. As we are mostly interested in the elemental ratios and relative abundances of the considered ices, lower ice fraction is an appropriate simplification. However, we note that dust dynamics can lead to accumulation of ices and ice mantles exceeding masses of silicate cores in some disc regions, as was shown previously in~\citet{2021ApJ...910..153M}.

Ice mantles also provide feedback on dust evolution. The model includes the effect of ices on fragmentation velocity $v_{\rm frag}$, which is the the maximum collision velocity leading to sticking instead of fragmentation. According to laboratory experiments, icy grains have higher $v_{\rm frag}$ than bare silicate grains by an order of magnitude \citep{2009ApJ...702.1490W,2015ApJ...798...34G}. In \citet{2021ApJ...910..153M} we used the values of fragmentation velocity $v_{\rm frag}=1.5$ and~15\,m~s$^{-1}$ for bare and icy grains, correspondingly. Here we follow \citet{2019ApJ...878..132O} and adopt lower values of $v_{\rm frag}=0.5$ and~5\,m~s$^{-1}$, which are more relevant for grains consisting of $\mu$m-sized monomers. These lower values of $v_{\rm frag}$ will lead to higher importance of fragmentation compared to \citet{2021ApJ...910..153M}. To determine if a dust grain should be considered icy or bare, we compare the local total surface density of all ices on grown dust divided by $\Sigma_{\rm d,gr}$ with the threshold value $K$, which is calculated as 
\begin{equation}
 K = \frac{3 a_{\rm ml}\rho_{\rm ice}}{\sqrt{a_*a_{\rm max}} \rho_{\rm s}},
 \label{eq:threshold}
 \end{equation}
i.e., an icy grain must have at least one monolayer of ice. Here, $a_{\rm ml}$ is the thickness of the ice monolayer estimated as the size of a water molecule $3 \times 10^{-8}$\,cm. The material density of ice $\rho_{\rm ice}=1$\,g\,cm$^{-3}$ and the mean radius of a grown grain is calculated as $\sqrt{a_*a_{\rm max}}$ for the power-law distribution with $p=3.5$.

\section{Results}
\label{sec:results}

To understand the distribution of the species, we first need to consider the global evolution of the disc and its structure. The distribution of volatiles and the C/O ratio is very sensitive to gas and dust substructures appearing in the disc. Variations in temperature and pressure lead to the complex shape and temporal evolution of the snowlines. The dependence of dust fragmentation velocity on the presence of ice mantles implies the feedback from the volatiles on dust and (through back-reaction) on gas.

Our modelling starts with the gravitational collapse of a flattened, slowly rotating molecular cloud. The protoplanetary disc is formed after the formation of the protostar, when the in-spiraling layers of the contracting cloud hit the centrifugal barrier near the inner edge of the sink cell, at a time instance depending on the initial core mass. The disc and the protostar are formed $\approx53$\,kyr after the beginning of the cloud collapse in model~M1 and at $\approx78$\,kyr in model~M2. If not stated otherwise, times are specified counting from the beginning of the simulation, e.g. the 100\,kyr time instance for model~M1 refers to a $\approx50$\,kyr old disc, as the first stage includes cloud collapse as well.

An important characteristic of young stellar objects is their variable accretion rate. Our modelling produces accretion bursts with the magnitude of $\sim100$\,$L_{\odot}$ occurring every $\sim10^{4}$\,years during the first hundred thousands years of disc evolution. These burst parameters are in line with the episodic accretion scenario \citep{1985ApJ...299..462H} and resemble the observed phenomenon of FU~Ori type stars \citep[see][for a review]{2014prpl.conf..387A}. The luminosity outbursts heat up the disc and typically shift the snowlines further away from the star. Although this effect is temporary, it can be reflected in the distributions of the volatiles, and leave long-term imprints in the observed dust properties~\citep{2022A&A...658A.191V}. The detailed analysis of the effect of such outbursts on the distribution of the elements is worthy of a separate study and lies beyond the scope of this paper.

\subsection{Dust and gas structures}
\label{sec:dustgasstructures}

\begin{figure*}
\includegraphics[width=\columnwidth]{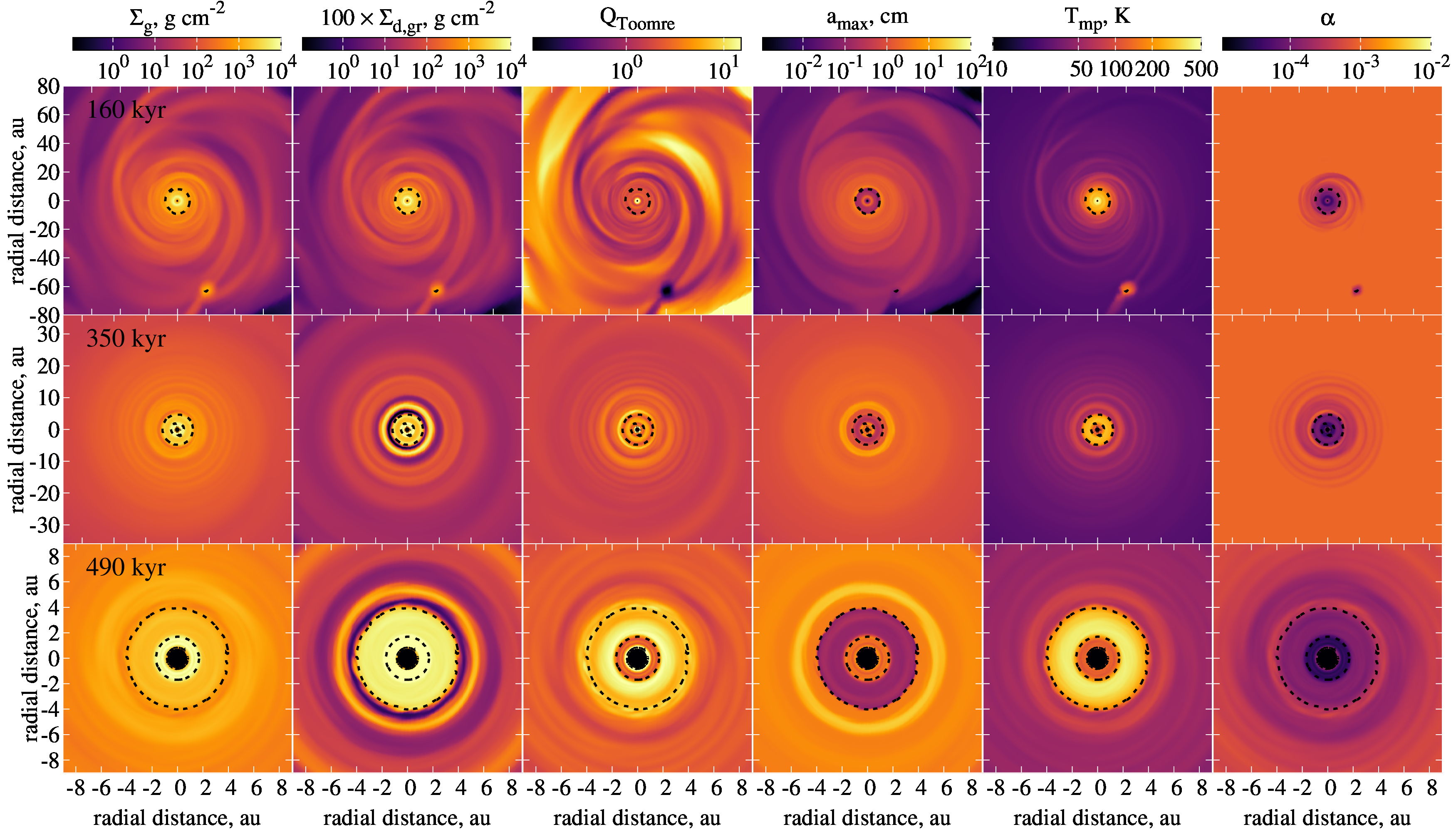}
    \caption{Surface density of gas and grown dust, Toomre $Q$-parameter, maximum dust radius, temperature and viscous $\alpha$-parameter in model~M1 at selected time moments: 160\,kyr,~$80\times80$\,au; 350\,kyr,~$35\times35$\,au; 490\,kyr,~$9\times9$\,au. The contours indicate the position of the water snowline. Note that at the panels with multiple water snowlines, water is frozen outside the outer line and inside an inner dust ring at $1-2$\,au.}
 \label{fig:maps1}
\end{figure*}

During the first hundred thousands years of evolution, protoplanetary discs change from highly asymmetric and dynamic objects to nearly axisymmetric and slowly evolving structures. Figure~\ref{fig:maps1} shows the examples of different gas and dust substructures that are present in the disc at different stages. The snapshots are shown for model~M1, they include a young disc (160\,kyr), an intermediate state (350\,kyr), and a more evolved and axisymmetric disc (490\,kyr). Each of the selected time instances represent some characteristic morphological features addressed below. In model~M2, similar structures appear, although sometimes at different evolution times. In this subsection, we consider model~M1 as an example and discuss these features, highlighting the properties of dust and gas substructures that are most relevant for the distribution of the volatiles.

The earliest phases of disc evolution are characterised by a large-scale spiral structure in both dust and gas, as well as episodic appearance of clumps. They are the result of gravitational instability (GI) in a massive disc, as in our modelling, the disc mass comprises more than $0.1$ of the stellar mass, which is roughly a threshold of the disc stability against GI~\citep{2013A&A...552A.129V,2016ARA&A..54..271K}. This is illustrated by the Toomre $Q$-parameter \citep{1964ApJ...139.1217T} in the third column of Figure~\ref{fig:maps1}: inside the spirals and the clump $Q<1$, which indicates the dominance of self-gravity over Keplerian sheer and gas density. The spiral structures become less prominent with time as the disc looses mass and the $Q$-parameter increases. However, spirals persist in the model throughout the disc evolution up to 0.5\,Myr. For example, at 350\,kyr, a very tight spiral is present in the gas, starting at the gas and dust ring at $\approx10$\,au. At 490\,kyr, the spiral pattern is weak and exists at $>10$\,au distances, which are not displayed in Figure~\ref{fig:maps1}. 

One- or two-armed grand design spirals are common $100-200$\,kyr after the disc formation in the models. The analogues of such spirals in the observed young protoplanetary discs around low-mass stars are found, for example, in Elias~2-27 or WaOph~6 \citep{2016Sci...353.1519P,2018ApJ...869L..43H}. It is not yet clear if the observed spirals are the result of GI or caused by a perturbation from a companion planet or a (sub)stellar object \citep{2017ApJ...839L..24M,2021A&A...654A..35B}. A spiral with multiple clumps produced by GI was recently observed in the disc around a FUor V960~Mon~\citep{2023ApJ...952L..17W}.

Another important feature of gas and dust spatial distributions is ring-like structures at various scales. The system of rings starts to form as early as 100\,kyr, and develops to the high-contrast multiple ring structure ($1-3$ orders of magnitude difference between surface densities in rings and gaps), which is evident in the middle and bottom rows of Figure~\ref{fig:maps1}. While gas rings are also common, the annual structures are more prominent in the dust surface density, as well as in dust size. Some of the dust rings correspond spatially to the gas rings, while others are barely reflected in the gas distribution. Overall, the difference between the gas and dust structures develops with time, as the result of dust growth and drift \citep{2014prpl.conf..339T}. Only some particular substructures, such as the ring between $1-2$\,au, are present in the distributions of both gas and dust components.

Another location where prominent rings form in both dust and gas is the water snowline. Here, the water snowline is defined as the location in the disc midplane where the amount of water in the gas equals to its amount in the ice (on both dust populations). There are multiple location in the disc where this happens. Water is frozen in most of the disc beyond $5-10$\,au, and we will refer to the furthest snowline dividing these outer frozen region from the inner one with the gas-phase water as a primary snowline. Generally, the primary snowline is roughly circular, but it can have asymmetries due to the spiral structure and an additional snowline may appear, e.g., around a gravitationally bound clump (upper rows in Figure~\ref{fig:maps1}). Besides, at $\approx260$\,kyr, another region rich in water ice appears in the inner disc, creating a pattern of double or triple water snowline at later times (middle and lower rows in Figure~\ref{fig:maps1}). We will refer to these inner additional snowlines as the secondary snowlines. They circumcise a cold and dense gas-dust ring that forms at $1-2$\,au. As the disc cools down with time, the primary snowline position moves from $\approx10$\,au distance at 160\,kyr, to $\approx6$\,au at 350\,kyr and $\approx4$\,au at 490\,kyr.

Snowlines are known to be associated with the enhancement of dust and volatiles \citep{1988Icar...75..146S,2004ApJ...614..490C,2017A&A...608A..92D,2021ApJ...910..153M}. Dust enhancement was also shown to affect the distribution of gas and its accretion rate through the disc \citep{2020A&A...635A.149G} by means of dust back reaction, which is also accounted for in our modelling. In our models, an about 2\,au wide dust ring is formed at the inner edge of the water snowline (at $\approx$10\,au) as soon as 50\,kyr after the disc formation. Grown dust grains drifting towards the star through the snowline lose their mantles, their fragmentation velocity drops, rendering them more vulnerable to fragmentation. Consequently, the grain maximum size $a_{\rm max}$ decreases, their drift towards the star slows down, which leads to the accumulation of grown dust, as well as small dust as a product of fragmentation. Increase of total dust density also leads to less efficient cooling and results in higher temperature inside the snowline (see the right panels in Figure~\ref{fig:maps1}). Later, at times $>200$\,kyr, several additional rings form outside the water snowline at distances up to 100\,au, the most notable one being $1-2$\,au exterior to the primary snowline. Dust is trapped inside the gas pressure maxima in these rings, which is a self-supporting phenomenon as the temperature also increases inside the dust ring due to high optical depth. 

These rings are worthy of attention in the context of possible planet formation. Dust surface density and size are higher in the rings, with $a_{\rm max}$ reaching centimetres, dust-to-gas ratio up to $0.1-0.2$, and Stokes number up to $0.05-0.1$. This could ease the development of the streaming instability, which typically requires pebble-size grains with $\mathrm{St}\gtrsim 0.01$ and dust-to-gas ratio $\gtrsim0.02$ \citep{2015A&A...579A..43C}. Multiple ring-like structures are commonly observed in protoplanetary discs at a range of ages and display a variety of examples, with different widths, contrasts and numbers of rings \citep[][and many others]{2018ApJ...869L..42H}. However, to directly compare the ring structures in the simulated dust surface density with the observed dust emission, require radiative transfer modelling is needed. Some of the observed ring structures could be produced by radial variation in dust size even in the absence of gaps in dust surface density \citep{2020MNRAS.499.5578A}.

The most prominent dust rings are located in the vicinity of the water snowline: the ring outside the primary snowline at $5-8$\,au (depending on the time) and the ring at $1-2$\,au, inside the primary snowline, which at later times also contains water ice and additional snowlines. The main effect of the snowlines is the change in fragmentation velocity between mantled and non-mantled grains: dust size sharply decreases by $\approx2$ orders of magnitude for the latter. Immediately outside the water snowline, the midplane temperature is lower, due to lower dust opacity and thus more efficient cooling. Turbulent $\alpha$ is on the contrary, higher, providing more efficient radial transport of matter. It leads to lowering the gas surface density in this gap, which in turn increases $\alpha$ (see Eq.~\eqref{eq:alpha}), creating the positive feedback and further deepening the gap. One of the possible mechanisms to create the initial decrease in the gas surface density is dust back-reaction, which can affect the inward flow of gas at the snowline. This effect was investigated by \citet{2020A&A...635A.149G} for different initial dust-to-gas ratios. The radial variation of $\alpha$ itself lead to the appearance of gas substructures \citep{2024MNRAS.tmp.1708T}. The combined effect of lower temperature and density creates a pressure minimum, which dust tends to avoid.

The dead zone, where $\alpha$-parameter values are lower than $10^{-3}$, includes the ice-free inner disc and has an approximately two times larger radial span than the iceless region. The distribution of $\alpha$-parameter is shown in the right column of Figure~\ref{fig:maps1}. Inside the primary water snowline, the values of $\alpha$ are the lowest due to high surface density of gas. The dead zone is not axisymmetric and reflects the spiral structures of the gas distribution, as it is sensitive to $\Sigma_{\rm g}$ (see Eq.~\eqref{eq:alpha}). The spiral arms of the dead zone span to $15-25$\,au at 160\,kyr, to $10-18$\,au at 350\,kyr and to $10-14$\,au at 490\,kyr. The comparison with the temperature distribution in Figure~\ref{fig:maps1} indicates that $\alpha$ and $T_{\rm mp}$ are anticorrelated, as higher viscosity provides faster accretion, hence lower surface densities and more efficient cooling. Similarly, a lot of dust accumulates in the dead zone, increasing opacity, which hampers cooling.

The icy dust ring at $1-2$\,au is especially interesting in the context of planet formation. In this ring, dust-to-gas ratio exceeds 0.1, and surface densities of both gas and dust are increased by more than an order of magnitude compared to the adjacent regions. Due to the ice mantles that protect dust from fragmentation, the dust size reaches several cm, close to the values behind the water snowline. When the ring is established, it is self-supporting, in a sense that without external perturbation (e.g. sharp change in accretion flow from the outer disc), it can be stable for a long time, over tens of kyr. 

The presence of water ice inside the dead zone was investigated by \citet{2023MNRAS.519L..10V}. They showed that in the discs around lower-mass stars, the turbulent heating in the inner disc can be low enough to allow the existence of ices. In our modelling the cooling of the inner region is assisted by the inner dust ring. The freeze-out has a positive feedback on dust growth due to higher $v_{\rm frag}$, which helps dust grow larger and further accumulate toward the pressure maximum in the ring.  So in our modelling ices coexist with a lot of grown dust in the dead zone. Such an icy inner region appears to be a promising place for the formation of the volatile rich planets.

\subsection{Distribution of volatiles}
\label{sec:volatiles_dist}

\begin{figure*}
\includegraphics[width=\columnwidth]{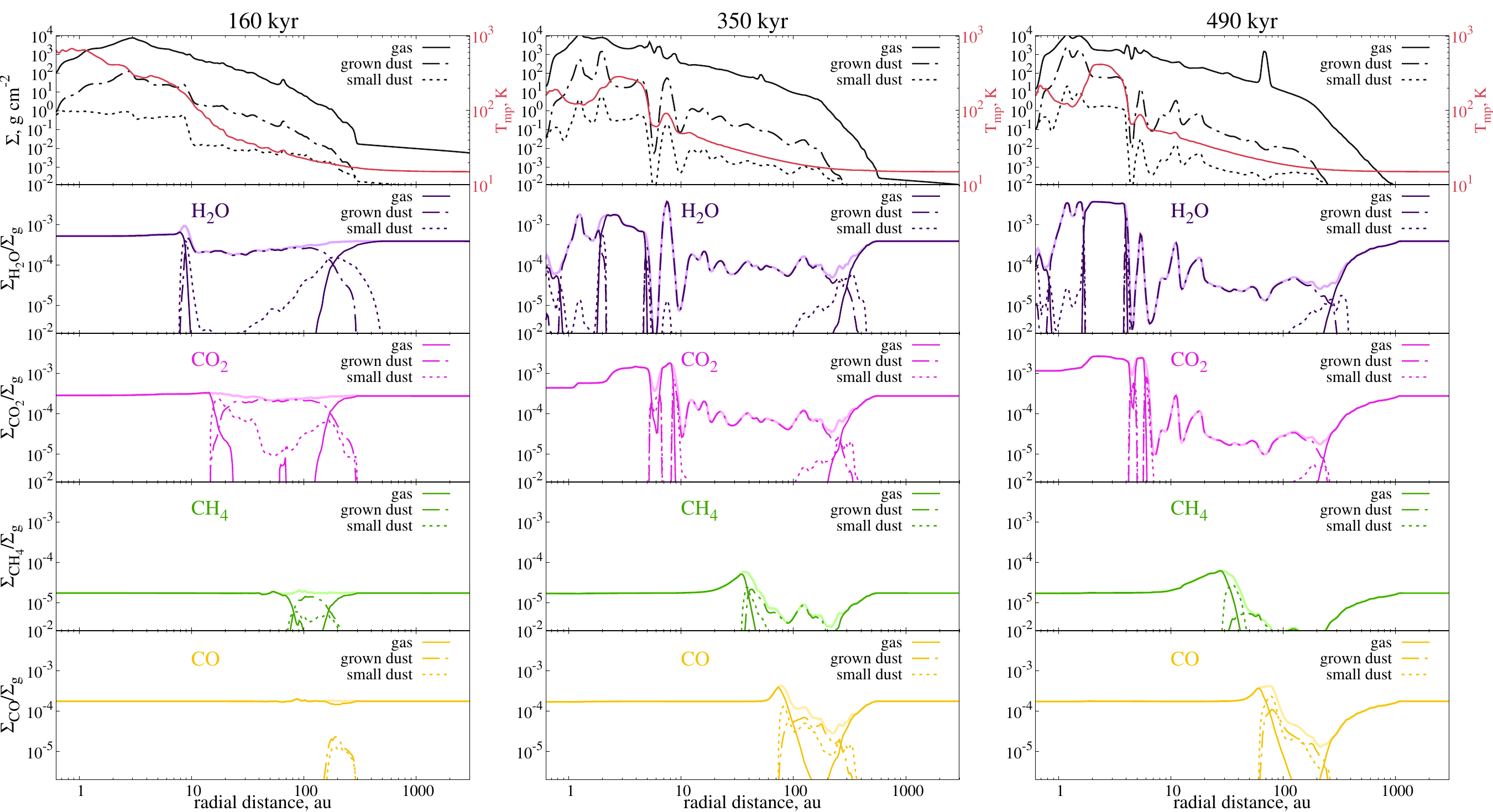}
 \caption{Radial distribution of azimuthally averaged surface densities of the volatiles in the gas and in the ice at various time instances in~M1 ($M_{\rm core}=0.66$\,$M_{\odot}$). Pale lines indicate the total surface density of species. The upper panels show surface densities of gas, small dust and grown dust, and the midplane temperature.}
 \label{fig:volatilesM1}
\end{figure*}

\begin{figure*}
\includegraphics[width=\columnwidth]{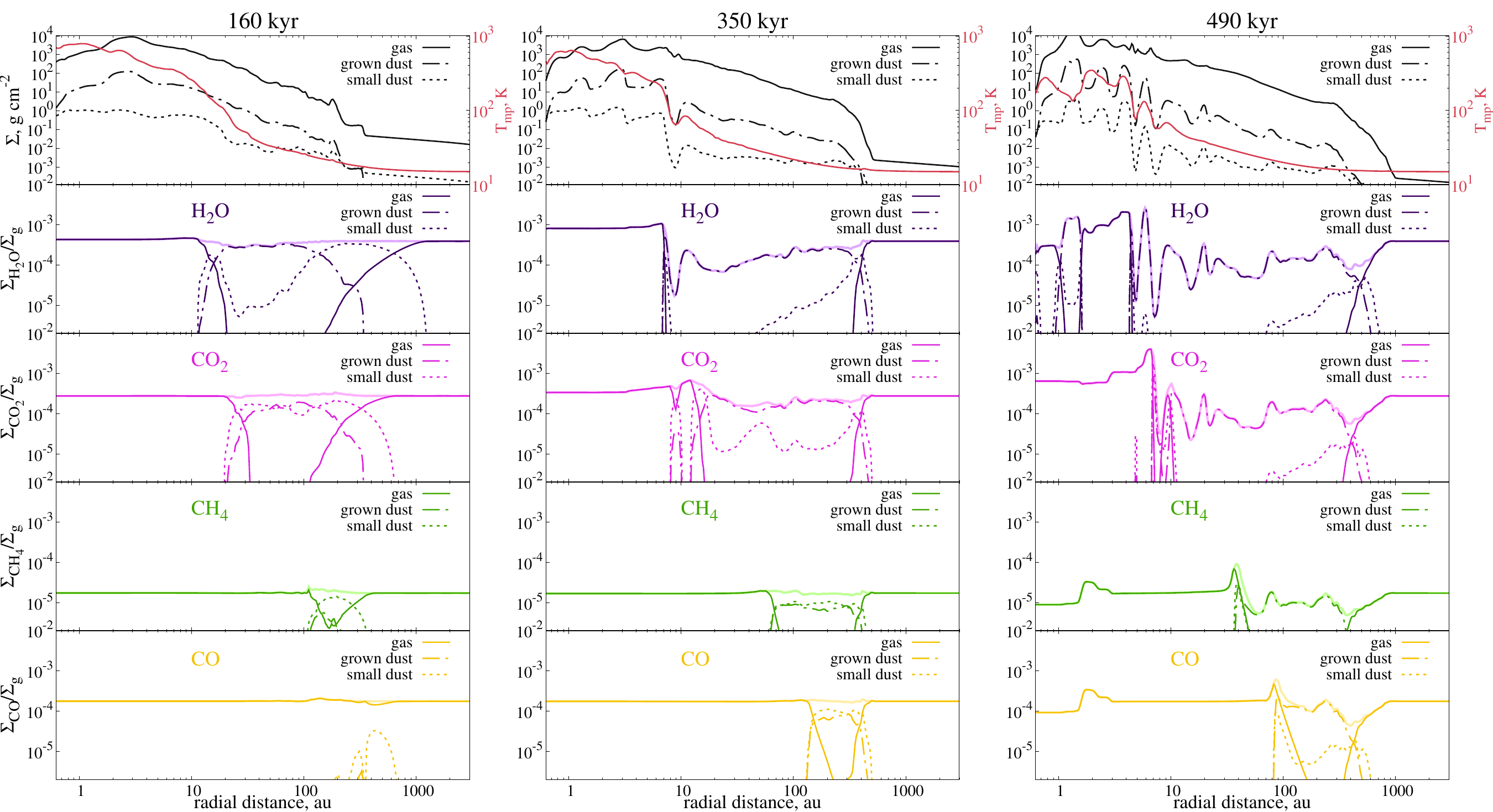}
 \caption{Same as Figure~\ref{fig:volatilesM2} but for model~M2 ($M_{\rm core}=1$\,$M_{\odot}$).}
 \label{fig:volatilesM2}
\end{figure*}

Even in the absence of chemical reactions, over the years of disc evolution the distribution of volatiles significantly changes compared to the initial one. This is the result of both phase transitions and dust growth and advection. Dust drift brings ices from the outer disc, enriching the inner disc with volatiles. Snowlines provide the conditions favouring accumulation of ices and gas-phase species. The formation of the established disc structures, such as dust and gas rings and spirals (see Section~\ref{sec:dustgasstructures}) leads to a complex pattern of intermittent snowlines. In this Section, we describe the main features of the molecular distributions and their implications for the composition of dust and gas at early stages of protoplanetary disc evolution.

Figures~\ref{fig:volatilesM1} and~\ref{fig:volatilesM2} show the azimuthally averaged radial profiles of the volatile species in models~M1 and~M2, respectively. As mentioned earlier, we define the snowline position as the location in the disc midplane where the amount of species in the gas equals to its amount in the ice (on both dust populations). This definition allows the snowline to have complex shape, characterised by different positions for each azimuthal direction. In the azimuthally averaged distributions shown in Figures~\ref{fig:volatilesM1} and~\ref{fig:volatilesM2}, the position would only be approximate. The slope of the species distribution in the vicinity of the snowline reflects the degree of the axial asymmetry in the disc. Flatter distributions appear as the contributions sum up from the snowlines, the radial position of which vary at different azimuthal angles~$\phi$. 

In our model setup, the volatiles can either have no snowline in the disc, or have two or more snowlines depending on the local conditions. Snowlines are absent for more volatile species (CO and CH$_4$) at earlier times or during bright luminosity outbursts, when the disc is too hot for them to freeze out. When there are two snowlines, the inner snowline in the disc is the one determined by thermal desorption. We refer to it as the primary snowline. The outer snowline is determined by photo-desorption, it lies in the embedding envelope outside of the body of the disc where optical depth is low. It must be noted that the gas-phase species outside this photo-snowline are vulnerable to photo-dissociation by the~UV radiation. This process is not explicitly included in our chemical model, but can be assessed as in \citet{2021ApJ...910..153M}.

More than two snowlines appear when the disc physical structure becomes more complex, mainly due to the presence of the ring-like structures. Species can freeze inside a cold dense dust ring, creating additional secondary snowlines, also governed by thermal desorption. The concepts of primary and secondary snowline is necessary for H$_2$O and CO$_2$, which have multiple snowlines at the later stages of disc evolution. For water, the inner icy region appears at $\approx1$\,au at 490\,kyr (see right column of Figure~\ref{fig:volatilesM2}) inside a dense dust ring. For CO$_2$, the ring that appears at 7\,au, outside the primary water snowline at 4\,au, creates the inverted thermal structure in the region with $T_{\rm mp}$ close to the typical CO$_2$ sublimation temperature of $70-90$\,K. Increase in $T_{\rm mp}$ in these ring is caused by higher optical depth and consequently lower cooling on the viscously heated midplane.

Apart from the snowlines, the distributions of volatiles in Figures~\ref{fig:volatilesM1} and~\ref{fig:volatilesM2} present local radial variations in all of the species components, including total abundance of the species. The initial total abundance is kept only in the envelope. As the matter is being redistributed, abundances of all volatiles in the inner disc grow. Particularly, the process responsible for this is dust drift. It brings the grown ice-covered grains to the inner disc regions, where their ice mantles are sublimated and no longer move with the drifting silicate dust cores. The effect is stronger for less volatile species H$_2$O and CO$_2$: their abundance in the gas grows by a factor of a few inside their primary snowlines. For CO and CH$_4$  the effect is weaker, because there is less grown dust outside their snowlines, and those snowlines are not very much established at earlier times. Thus their abundances inside the snowline only grow by a factor of unity, except for the immediate vicinity of the snowline. Moreover, both CO and CH$_4$ have a bump in the gas-phase distribution just outside the dust ring at 1\,au, that is absent in H$_2$O and CO$_2$. At the inner disc edge the abundances of CO and CH$_4$ in the gas are lower than the initial value.

The abundance enhancements appear most notably at the snowlines, with local bumps in all three phases at later times (after $\approx350$\,kyr). They are produced by the combination of the dust radial drift and the azimuthal oscillations of dust and gas radial velocity, described in more detail in \citet{2021ApJ...910..153M} and Molyarova et~al., in prep. By 500\,kyr, the surface density of gas-phase H$_2$O exceeds the initial value by a factor of 5, of CO$_2$ -- by a factor of 7, of CH$_4$ -- by 3.5 and of CO -- by 2.5. Similar accumulation powered by turbulent diffusion was previously studied for CO molecule in axially symmetric model setup \citep{2017A&A...600A.140S,2018ApJ...864...78K}. Their results indicated similar enhancement in the gas phase by a factor of a few. In our non-axisymmetric approach, diffusion is not a necessary requirement, and the necessary transport is rather provided by azimuthal variations of radial velocity induced by disc self-gravity (Molyarova et~al., in prep).

Distributions of ices on small and grown dust are different as they are affected by dust growth and drift. In general, there is more grown dust than small by mass, and in most of the disc there is more ices on grown dust, particularly at later times. However, in the outer disc and at the earlier times, ices on small grains dominate. As initially all ices are on small dust, it seems inevitable that they will gradually move to the grown grains as small dust coagulates and turns into grown grains. However, near all the snowlines, amount of ices on small dust increases due to the effect of the spiral pattern. Ice abundances are determined by episodic sublimation and freeze-out in the warm spiral arms of the complex density and temperature pattern \citep[see Section 3.3.2. in][]{2021ApJ...910..153M}. As adsorption preferably happens to the smaller grains due to their larger total surface area, there is much more ices on small dust in the wake of the spiral arms. This concerns the photo-desorption snowlines as well.

\subsection{C/O ratios}
\label{sec:C2O_ratio}

\begin{figure*}
\includegraphics[width=0.49\columnwidth]{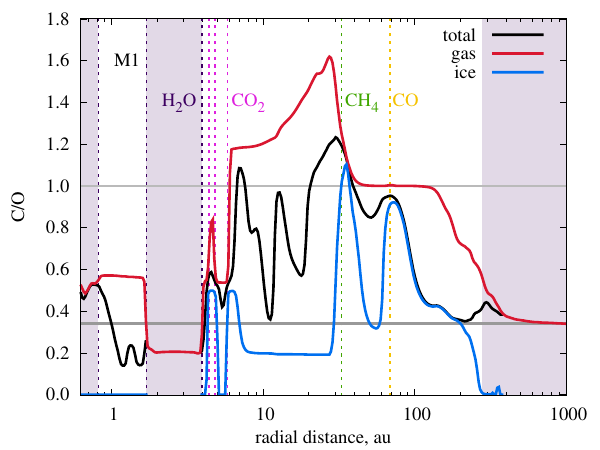}
\includegraphics[width=0.49\columnwidth]{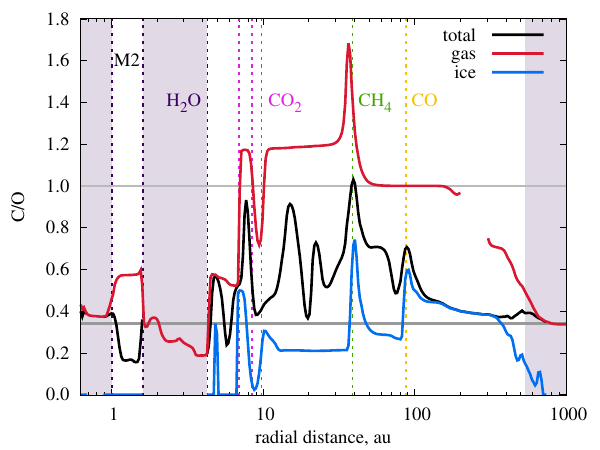}
 \caption{Radial profiles of the C/O ratio at 490\,kyr in models~M1 (left) and~M2 (right). The plots show C/O in total (black), in the gas (red), and in the ice (blue). The C/O ratio in the ice (gas) is only shown for radial distances where the mass of the volatiles in the ice (gas) is larger than 0.1\% of the total mass of the volatiles in the gas (ice). The grey horizontal line indicates the baseline $\mathrm{C/O}=0.34$. Positions of the snowlines are highlighted by vertical dashed lines. The regions where water (thus all other species) is in the gas are shaded with purple.}
 \label{fig:c2o_radial}
\end{figure*}

Here we consider the relative amount of carbon and oxygen in the gas, on the surface of small and grown grains, and in total for all for all phases and disc locations. C/O ratio is seen as a perspective tracer of planet formation mechanisms \citep[see][]{2011ApJ...743L..16O,2015A&A...574A.138T}. Therefore, we are especially interested in the regions of the disc and the volatile phase component where the C/O ratio declines from the total initial value (see below). Particularly, we are interested in C/O ratio noticeably above the initial value, as it was observed in atmospheres of exoplanets and suggests the formation of these planets in similarly carbon-enriched environments \citep{2009ApJ...704.1616S,2011Natur.469...64M,2013ApJ...763...25M,2021AJ....162...99F}. 

The C/O ratio is calculated as the total number of carbon atoms contained in the molecules, divided by the total number of oxygen atoms in the molecules. This calculation can be done separately for the species contained in the gas phase, species on dust surface, or for the species in all phases. Thus, we calculate the C/O ratio in the gas, in the ice, and total C/O, respectively. For the ice species, we include both ices on grown and small dust grains.In some disc regions, the amount of carbon and oxygen in in a particular phase is very low, e.g. in the ice phase inside the water snowline, where all the molecules are in the gas. For such cases, C/O ratio would not be representative of the chemical composition, so we exclude the corresponding computational cells from the consideration. We only display the C/O ratio if the mass of the volatiles in the phase is higher than $10^{-3}$ of the total mass of volatiles at a given location.

For the adopted molecular abundances based on \citet{2011ApJ...740..109O} and also used by \citet{2018A&A...613A..14E}, the baseline value is $\mathrm{C/O}=0.34$, which is in line with the gas-phase abundances in the ISM~\citep[$\approx0.36$,][]{2008ApJ...688L.103P}. This value is different from the typical solar value of $\approx0.5$~\citep{2008ApJ...688L.103P}, which is commonly used as a reference \citep[e.g.,][]{2011ApJ...743L..16O,2018A&A...617A..28S}. First, local galactic abundances changed since the Solar system formation 4.6\,billion years ago \citep[see, e.g., Appendix~A in][for the comparison]{2024arXiv240612037B}. Second, the difference also stems from inclusion or non-inclusion of the dust component. The stellar atmospheric abundances comprise all the elements present in the medium where the star formed, while the elements available for the volatiles are only a fraction of that. The values of the elemental abundances can be determined separately for the volatile (gas and ices) and the refractory (rocky dust grain cores) components of the ISM \citep{2021ApJ...906...73H}. Here, we do not take into account carbon and oxygen from the rocky cores of dust grains. Due to the model assumptions, the initial ice-to-rock mass ratio in our model is quite low \citep[0.08 compared to $2-4$ in][]{2014prpl.conf..363P}. Therefore, adding the elements from solid dust grains to those contained in ices would be misleading, as the former would dominate in the resulting C/O ratio. In this work, we concentrate on considering the C/O ratios of the volatile component (ice and gas), and compare them with the initial value of~0.34.

Figure~\ref{fig:c2o_radial} shows radial profiles of the C/O ratios averaged over the azimuthal angle $\phi$ by  the end of the simulations, at 490\,kyr. The key values of C/O ratio relevant in the context of planet formation are the initial value of $0.34$, motivated by the comparison with the initial abundances, and $1.0$, which is a boarder between carbon- and oxygen-dominated chemical regimes in the ISM and (exo)planetary atmospheres~\citep[see, e.g.,][]{2005pcim.book.....T,2005ApJ...632.1122S,2012ApJ...758...36M}.

By the age of 490\,yr, the C/O ratio in both models is significantly different from the initial value. This concerns the total C/O ratio, as well as the C/O ratio in the gas and in the ice. Let us analyse the distributions of C/O ratios that we can see from Figure~\ref{fig:c2o_radial}, moving inwards from the envelope to the centre. The main features of the C/O distributions are the following:
\begin{itemize}
    \item the C/O ratio in the envelope is close to the initial value $0.34$ in  the gas and in total, and it grows up to unity from the envelope to the CO snowline in the disc;
    \item around the CO snowline, the C/O approaches unity;
    \item between CO$_2$ and CH$_4$ snowlines the C/O in the gas is $>1$, and the C/O in the ice is below the initial;
    \item at the distances of tens of au, there are variations in the total C/O ratio that are not connected with any snowlines, or variations in gas- or ice-phase C/O;
    \item ice-phase C/O ratios peak around all snowlines except water;
    \item inside the primary water snowline is the region rich in volatiles, with both gas- and ice-phase C/O are the lowest in the disc.
\end{itemize}
As expected, the key changes in the C/O ratio distribution are associated with the positions of the snowlines. There are two main processes. First, the freeze-out and desorption at the snowlines transfer the elements between phases, altering the C/O in the gas and in the ice. Second, the snowlines favour the accumulation of the respective volatiles both in the gas and in the ice, as was discussed in Section~\ref{sec:volatiles_dist}, pumping up the amount of both components and altering their proportion. Besides the snowlines, radial drift of grown grains~\citep{1977MNRAS.180...57W} transports the volatiles inwards. We discuss below which processes are responsible for the formation of the listed features.

\underline{\textit{C/O in the outer disc.}} In the surrounding envelope, outside the disc, all volatiles are in the gas due to photo-desorption, and C/O in the gas is close to the initial value of 0.34. Around the CO snowline and beyond, the C/O in the gas is close to unity, and in the gas C/O is higher that the initial, which requires explanation. The region beyond the CO snowline is usually described as the place where all species are frozen, thus having a stellar C/O ratio in the solid phase and practically no carbon or oxygen in the gas \citep[e.g.,][]{2011ApJ...743L..16O,2019AJ....158..194O,2020A&A...640A.131M}. In our simulations, only in model~M2 there is a region where less than $10^{-3}$ of C and O is in the gas, and  in model~M1 such region is absent. Due to the asymmetric spiral structure that persists even at 490\,kyr, even though most of CO is frozen beyond the snowline, there is a significant ($>10^{-3}$) fraction of it in the gas. Additionally, the position of CO snowline itself is significantly affected by the dust drift, as it declines from the equilibrium between adsorption and desorption.The indistinctness of the CO snowline also helps CO to persist in the outer regions: all other species are ultimately frozen, so they are efficiently carried away to the inner disc via radial drift, while this works worse for only partially frozen CO. It creates relative overabundance of CO in the outer disc, elevating the total C/O ratio and later the ice-phase C/O ratio, when the preserved CO freezes out. Between the CO ice line and the envelope, there is a gradient of C/O in the gas  due to photo-desorption of the ices. The last molecule to be photo-desorbed is water, which returns oxygen to the gas phase at the farthest radial distance.

\underline{\textit{$C/O\approx1$ around the CO snowline.}} Beyond the CH$_4$ snowline, CO dominates the composition of volatiles in the gas phase, leading to the gas-phase C/O close to unity, the value characteristic of the CO molecule. At the CO snowline, the CO dominates the ice-phase composition as well. While dust drift substantially lowered the abundances of CO$_2$ and H$_2$O by 490\,kyr in these regions, CO ice accumulated at the snowline. This leads to the ice-phase C/O closer to 1, too, making the vicinity of the CO snowline a region where total amounts of C and O are similar.

\underline{\textit{High C/O in the gas, low C/O in the ice.}} In the disc regions beyond the primary CO$_2$ snowline, only CO and CH$_4$ are in the gas, meaning the dominance of carbon and C/O~$>1$. Consequently, the C/O in the ice is generally lower, around $0.2$, as the ices are mainly H$_2$O and CO$_2$ rich in oxygen. This is consistent with the classical step-like picture of \citet{2011ApJ...743L..16O}, with the addition of carbon-rich methane allowing the $\mathrm{C/O}>1$.
The midplane C/O ratios we simulate are difficult to directly compare with observations, which mostly trace the molecular layer above the disc midplane. High C/O ratios in the gas are indeed observed in many protoplanetary discs \citep[e.g.][]{2019A&A...631A..69M}, but they are considered to be a natural consequence of dust settling \citep{2018ApJ...864...78K,2020ApJ...899..134K}. Dust inward drift could also enhance this effect in the outer disc regions, creating the radial gradient of total C/O ratio in the disc. Observations of CS and SO emission coming from close to the midplane layers potentially indicate the presence of such radial gradient of the gas-phase C/O in the PDS~70 disc \citep{2024arXiv240706272R}.

\underline{\textit{Variations of total C/O.}} Throughout the disc, there are sharp changes in total C/O ratio not connected with any snowline. Most noticeable are the variations between CO$_2$ and CH$_4$ snowlines, where gas-phase and ice-phase C/O ratios are stable. These variations are associated with the disc substructures, particularly with the dense dust rings described in Section~\ref{sec:dustgasstructures}. The total C/O changes due to radial variations in dust-to-gas ratio: when it is higher, the total C/O is closer to the ice-phase C/O, and vice versa. Inside the dust-rich rings, H$_2$O and CO$_2$ ices are abundant, due to high surface density of dust relative to gas. At the same time, CO and CH$_4$ in the gas phase have similar surface densities inside dust rings and between them. Thus, in the dust rings, CO$_2$ and water are overabundant, leading to lower total C/O ratio. We consider this effect in more detail in Section~\ref{sec:2d_c2o}. Variations of the total C/O ratio due to dust substructures are also present in the cold ring at $\approx1$\,au.

\underline{\textit{C/O peaks at the snowlines.}} There are peaks of the C/O ratio in the ice right outside the snowlines of CO$_2$, CH$_4$ and CO, produced by the accumulation of the respective ices (see Figures~\ref{fig:volatilesM1} and~\ref{fig:volatilesM2}). In M2~model, the amount of CH$_4$ and CO ices at their respective snowlines becomes comparable  with or even larger than those of CO$_2$ and H$_2$O (compare the right panels of Figure~\ref{fig:volatilesM2}), leading to C/O in the ice $\approx0.6-0.9$. In model~M1, the accumulation is more prominent, so the C/O ratio in the ice approaches unity at CO snowline and $>1$ at the methane snowline. These peaks distort the pattern of generally low C/O ratio in the ice and preset additional regions where carbon-rich planetesimals could be formed, and carbon-rich pebbles could be accreted onto forming protoplanets. At the snowlines of H$_2$O and CO$_2$, the C/O in the ice approaches the C/O ratios of these molecules, 0 and 0.5, respectively.

\underline{\textit{Lower C/O inside the water snowline.}} Inside the primary water snowline, the C/O ratio is generally lower than outside of it, close to the initial 0.34. Contrary to the outer regions depleted of ices due to the radial drift, the disc parts inside and around the water snowline are enriched in volatiles, and particularly of oxygen-rich water and CO$_2$.
Total and gas-phase C/O ratios vary from $\approx0.2$ to $0.6$, as the secondary water snowlines add more substructure to the C/O distribution. The regions where the only ice is water and the ice-phase $\mathrm{C/O}=0$ are the inner cold dust ring at 1\,au and the narrow annuli between water and CO$_2$ snowlines. The enrichment of the inner disc regions with oxygen as a result of dust radial drift is suggested by the resolved observations of molecules in protoplanetary discs \citep{2020ApJ...903..124B}.

These characteristic features of the C/O distributions are similar in the two presented models. The main difference is the radial distances where the borders between the zones are located; they are closer to the star in the less massive and thus colder model~M1. This is mainly due to slightly different masses of the central star accumulated throughout 490\,kyr of non-identical protostellar accretion history, which lead to different luminosity and thermal structure (see upper panels of Figure~\ref{fig:c2o_m1_m2}). Particularly, the stellar masses and luminosities at this time instance are: $1.07$\,$L_{\odot}$ and $0.34$\,$M_{\odot}$~(for~M1), 1.89\,$L_{\odot}$ and 0.58\,$M_{\odot}$~(for~M2). The less massive model~M1 demonstrates overall higher C/O ratio in both phases.

\subsection{Evolution of the snowlines and the C/O ratios}
\label{sec:snowline_evolution}

\begin{figure*}
\includegraphics[width=0.49\columnwidth]{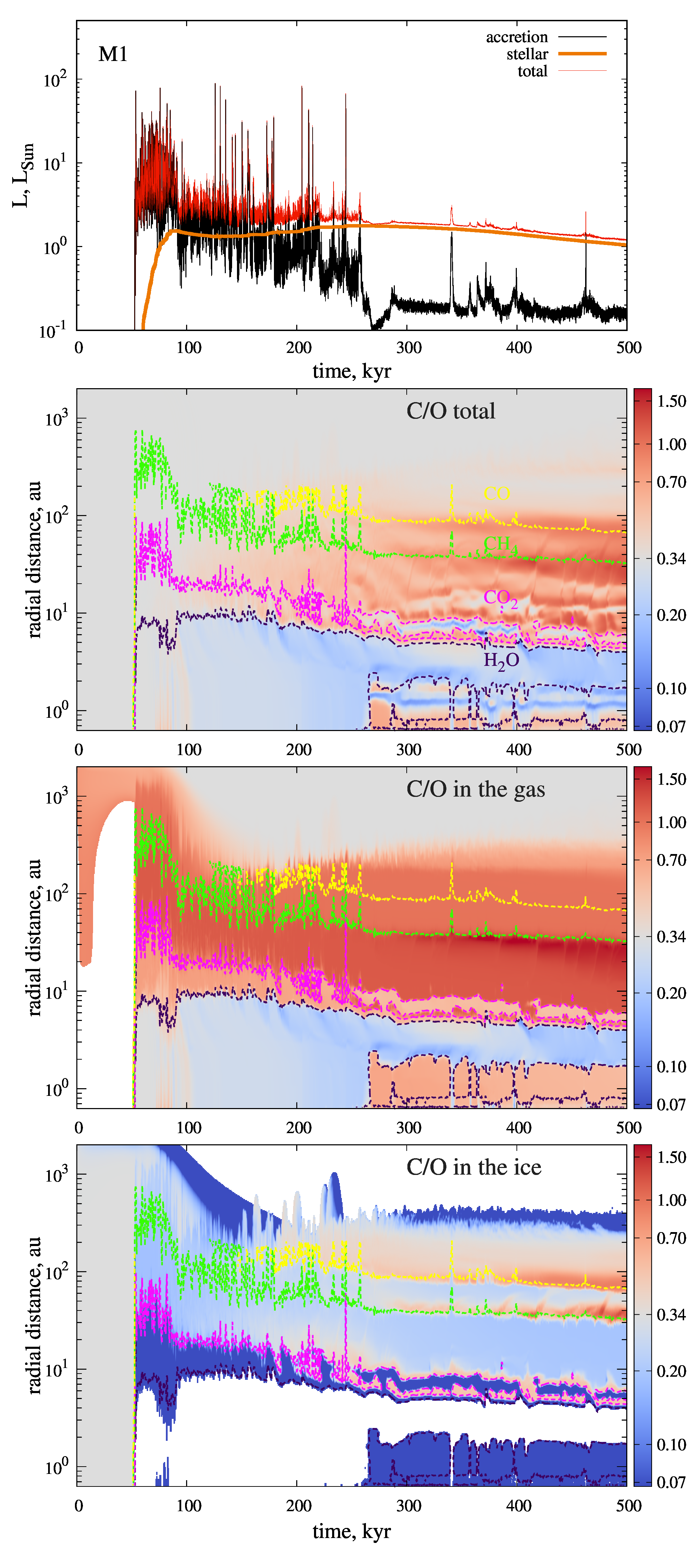}
\includegraphics[width=0.49\columnwidth]{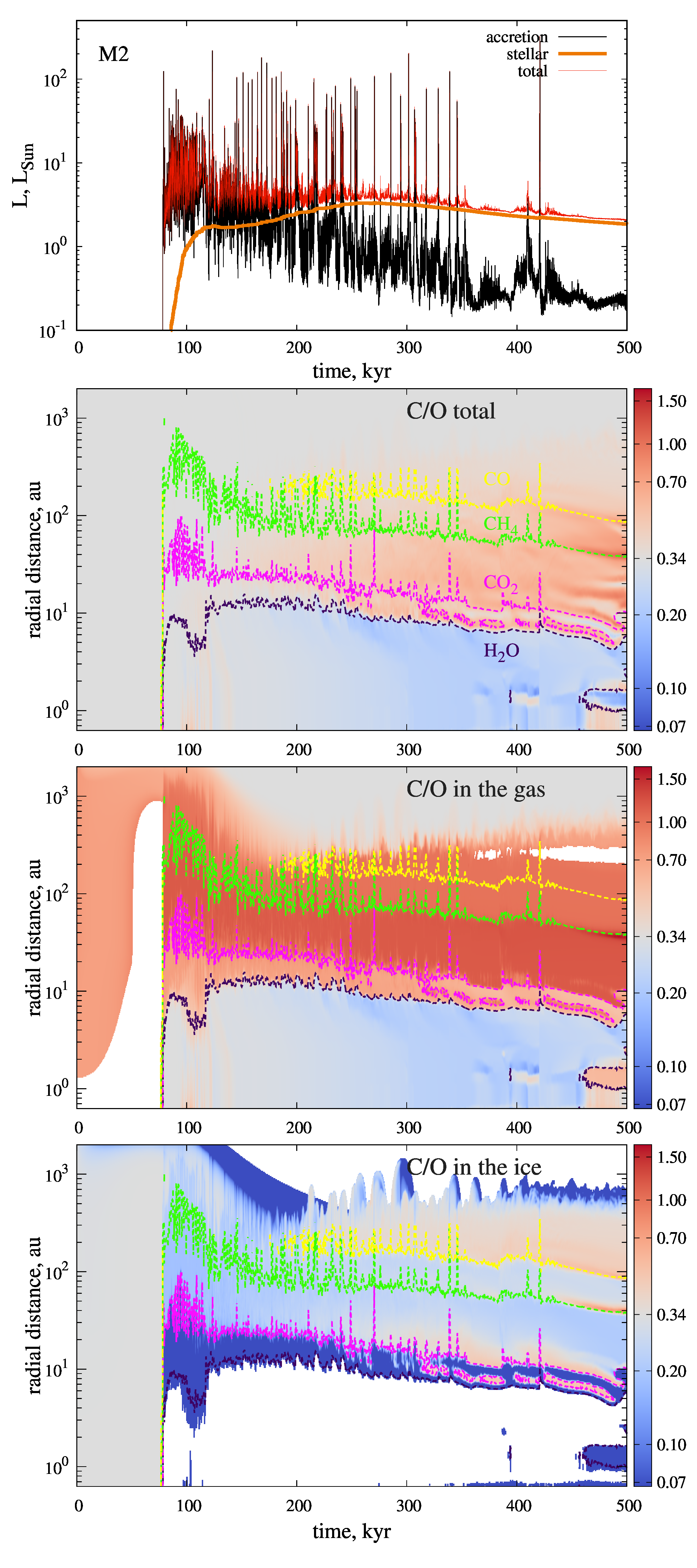}
 \caption{Evolution of central source luminosity and C/O ratio in models~M1 ($M_{\rm core}=0.66 M_{\odot}$, left) and~M2 ($M_{\rm core}=1 M_{\odot}$, right). The upper panels show stellar and accretion luminosity  depending on time. Below are, successively, total C/O ratio, C/O in the gas, and C/O in the ice, depending on time. The C/O values above and below the initial value of 0.34 are coloured in shades of red and blue, respectively. The regions with low abundances of both carbon and oxygen, either in the gas or ice phases, are shown in white. Coloured contours correspond to the positions of the snowlines. Photo-dissociation snowlines are not shown.}
 \label{fig:c2o_m1_m2}
\end{figure*}

The positions of the snowlines are crucial for the values of C/O ratios, both because of the direct change through the phase transitions and the associated accumulation of volatiles. They depend on the local gas and dust properties, particularly on temperature. They can also be shifted inward due to dust drift \citep{2015ApJ...815..109P}. Snowlines evolve as the disc structure changes with time. In this Section, we consider the co-evolution of the snowlines and the C/O ratios and discuss the mechanisms of  species redistribution over the disc.

The temporal evolution of the azimuthally averaged C/O ratios and the equilibrium positions of the snowlines is shown in Figure~\ref{fig:c2o_m1_m2}. It is evident from Figure~\ref{fig:c2o_m1_m2} that the C/O ratio indeed follows the snowlines, particularly inside $\approx10$\,au. The C/O structure changes throughout the disc evolution, and some key features only appear at later times. 

One of the key factors affecting the disc thermal structure is the luminosity of the central source. In the upper panels of~Figure~\ref{fig:c2o_m1_m2}, we show the evolution of total luminosity, which directly affects the positions of the snowlines. The luminosity is the sum of stellar and accretion components, coming from the protorstar itself and gravitational potential energy of the accreted matter. The stellar luminosity gradually decreases as the protostar becomes more compact. Accretion luminosity depends on the accretion rate, which is highly variable as a result of magnetorotational and gravitational instabilities in the disc \citep{2019ApJ...882...96K,2020ApJ...895...41K,2020A&A...644A..74V}. The simulated episodic luminosity outbursts are similar to those occurring in the observed YSOs \citep{2018ApJ...861..145C}, with their amplitudes of tens and hundreds $L_{\odot}$. In the more massive model~M2, the outbursts are more frequent, brighter and occur until later times. The reasons behind this difference is the massive disc being more prone to both MRI and GI, which needs to be investigated in more detail in a separate study.

Snowlines of the least volatile of the considered species, H$_2$O and CO$_2$, exist in the model since the earliest phases of the disc formation. Even during bright luminosity outbursts ($\sim100$\,$L_{\odot}$) they do not disappear, but move farther away from the star. During the first $\approx50$\,kyr the disc is spreading out, it is highly asymmetric and dynamic, so the snowline positions oscillate. Later on, the disc generally cools down, and the snowlines gradually move toward the star (except during the outbursts). In model~M2, between 130 and 500\,kyr, the primary water snowline moves from~12 to $4.5$\,au; the primary CO$_2$ snowline moves from 23 to 7\,au. In model~M1, the water snowline moves from~9 to $4$\,au, and the CO$_2$ snowline moves from 17 to 4.5\,au.

Despite a factor of two different binding energies of H$_2$O and CO$_2$ (5770 and 2360\,K, respectively), the locations of their primary snowlines do not differ much due to the steep radial temperature gradient around these distances (see upper panels of Figures~\ref{fig:volatilesM2} and~\ref{fig:volatilesM1}). Inside $10-20$\,au, $T_{\rm mp}$ is determined by heating mechanisms other than external irradiation: viscous heating, gas work (PdV heating), heating by shocks and energy transport with advection. This also means that the water snowline is less sensitive to the level of irradiative heating, thus only slightly affected by the luminosity outbursts. The temperature change is particularly sharp at the water snowline(s), with the absolute value of the approximated power-law slope of $1.5-6$. High surface density of dust in water-ice-free regions makes cooling less efficient and leads to locking up the produced heat and consequently higher temperatures. Besides, both these species have multiple snowlines due to the formation of ring-like substructures with  the conditions close to the borderline between their frozen and gaseous state. These additional snowlines also affect the C/O distributions.

Methane and CO are the more volatile species in our model. They either have zero or two snowlines in the disc. The snowlines are absent during the outbursts brighter than $\approx100$\,$L_{\odot}$ for methane and $\approx200$\,$L_{\odot}$ for CO. Until approximately $200-250$\,kyr, there is no established CO snowline inside the disc. For example, at 160\,kyr (see lower left panel in Figures~\ref{fig:volatilesM2} and~\ref{fig:volatilesM1}) there is CO ice on both small and grown dust, but their amount is an order of magnitude lower than that of the gas. Additionally, most of this ice is located at the outer disc edge, where gas and dust surface densities sharply drop. Similar distribution appears for CO and CH$_4$ during bright outbursts: their ices are present in some disc regions, but due to non-axisymmetric disc structure, they do not dominate in the averaged profiles, and there is no common snowline for the whole disc. There are no secondary snowlines of CO or CH$_4$, because there is no prominent gas and dust structures in the outer disc where these species are frozen.

Snowlines divide the disc into several zones with different characteristic C/O ratios. However, the chemical composition and C/O ratios in these zones change with time. One of the distinct zones is the region where water is not frozen, shaded with purple in Figure~\ref{fig:c2o_radial} and circumscribed by the dark purple dotted line in Figure~\ref{fig:c2o_m1_m2}. In this zone, all the species are in the gas phase, so the C/O ratio is initially close to 0.34. However, as the disc evolves, dust brings more volatiles from the outer disc. Particularly, the abundance of water grows most significantly, making C/O in the gas decrease with time. This happens because the main mechanism of the transport of the volatiles is dust radial drift, which works best in the regions where grown grains can sustain their mantles. \citet{2020ApJ...903..124B} suggest that dust growth and drift can be responsible for the observed anticorrelation between disc radius and H$_2$O emission, implying that the inner regions of small discs with are enriched in water brought by efficiently drifting grains. Water is the least volatile species in our model, it is frozen in the largest part of the disc, thus its distribution is most strongly affected by the dust drift. At later times, this zone is divided into two, when a cold dense dust ring forms at $1-2$\,au, which happens at $\approx460$\,kyr in model~M2 and at $\approx270$\,kyr in model~M1. Interior to the ring, water abundance in the gas is around an order of magnitude lower than outside of it in both models. This happens because the inward flow of gas-phase H$_2$O is ``blocked'' by the cold ring where it freezes and accumulates with the grown grains in the pressure maximum. So the C/O ratio in the gas at $r\lessapprox2$\,au is determined by CO$_2$ and thus close to 0.5. 

The C/O ratio in the ice is not defined in the envelope, where all ices are photo-desorbed, and in the warmest inner disc, where even water is in the gas. In disc regions where only water is frozen, the C/O ratio in the ice is~zero. Before $200-250$\,kyr, when the disc cools down enough for CO to freeze out, the C/O ratio in the ice in the rest of the disc is close to 0.2. It is determined by CO$_2$ and H$_2$O ices, with a small contribution from the low-abundance CH$_4$. After that, when CO freezes out and CO snowline appears, the C/O ratio in the outer disc region becomes close to the initial value both in the ice and in total. This region beyond the CO snowline is frequently referred to in relation to giant exoplanets with stellar C/O ratios \citep[e.g.,][]{2020A&A...640A.131M,2019AJ....158..194O,2021A&A...651L...2O}. This region would be a perfect location for planets to accrete pebbles covered with icy mantles with the primordial elemental ratio, which would directly become part of the planetary atmosphere. However, pebbles are not initially present in the disc, and their existence in these outer regions is not guaranteed. Pebbles are dust grains large enough to move relative to gas \cite[e.g.,][]{2012A&A...544A..32L,2019ApJ...874...36L}, and a fraction of the grown dust in our modelling can be classified as pebbles.  The properties of pebbles and composition of their ice mantles in the same setting of the FEOSAD model were studied by \citet{2024MNRAS.530.2731T}. They show that pebbles appear in the disc as early as 50\,kyr after its formation, and exist in a wide  region of the disc. This partially includes the region beyond the CO snowline, but only the area of CO enhancement, where CO dominates in the ice composition and thus the C/O ratio in the ice is close to unity (see lower panels in their Figure~3).

As shown by \citet{2024MNRAS.530.2731T}, ices on pebbles are dominated by H$_2$O and CO$_2$. In this case, relatively high values of the C/O in the ice ($\approx0.5$) correspond to the regions where there is more CO$_2$, i.e. around the CO$_2$ snowlines. This is also the region where a prominent dust ring forms under the influence of the primary water snowline. It is characterised by accumulation of CO$_2$ and to lesser degree H$_2$O ice, as well as vapours, and relatively high amount of grown dust in the ring. The dust ring is situated between the two regions with C/O~$\approx0.34$ in the ice. It presents another favourable location for accreting the ice content with close-to-initial C/O ratio.

The total C/O ratio in the disc also changed significantly from the initial value due to dust drift that redistributes the ices. The matter becomes more carbon-rich as the grains bring CO and CH$_4$ from the outer disc parts. 
This effect was previously investigated  by  \citet{2017A&A...600A.140S} and \citet{2018ApJ...864...78K} in their modelling of CO dynamics and dust evolution. However, as was shown by \citet{2018ApJ...864...78K,2020ApJ...899..134K}, vertical settling of grown grains towards the midplane is responsible for depleting the upper layers in the outer disc of gas-phase oxygen, which cannot be captured within our thin-disc modelling.
The panels in the second row of Figure~\ref{fig:c2o_m1_m2} demonstrate strong enhancement of total C/O ratio in the intermediate disc regions. In model~M2, the total C/O ratio between $10-100$\,au becomes $\approx0.7$ after $\sim300$\,kyr, which is two times higher than the initial value. At the snowlines of CH$_4$ and CO$_2$ it approaches unity, mostly due to the accumulation of these species in the gas. In model~M1, this process begins $\approx200$\,kyr earlier and consequently leads to even higher total C/O ratio. At $400-500$\,kyr, most of the disc between~5 and~$100$\,au has total C/O~$\gtrsim1$ in model~M2, which demonstrates the powerful impact of dust drift.

A distinctive feature of the produced C/O distributions is the peaks of total and ice-phase C/O ratios around the CO and CH$_4$ snowlines. In model~M2, the increase of the C/O ratios becomes noticeable only after $\approx 450$\,kyr, while in model~M1 it starts to form around $\approx 300$\,kyr. Initial abundances of CH$_4$ and CO are lower than that of CO$_2$ and H$_2$O. As these species accumulate at their snowlines, the abundances become comparable, leading to C/O in the ice $\approx 0.6-0.8$ in model~M2 and up to 1 in model~M1. Similar accumulation is seen around CO$_2$ snowline, but unlike CO and CH$_4$ snowlines, it is also connected to the interaction with the dust ring structure and is considered in more detail in Section~\ref{sec:2d_c2o}.

\subsection{Two-dimentional structure}
\label{sec:2d_c2o}

\begin{figure*}
\includegraphics[width=0.49\columnwidth]{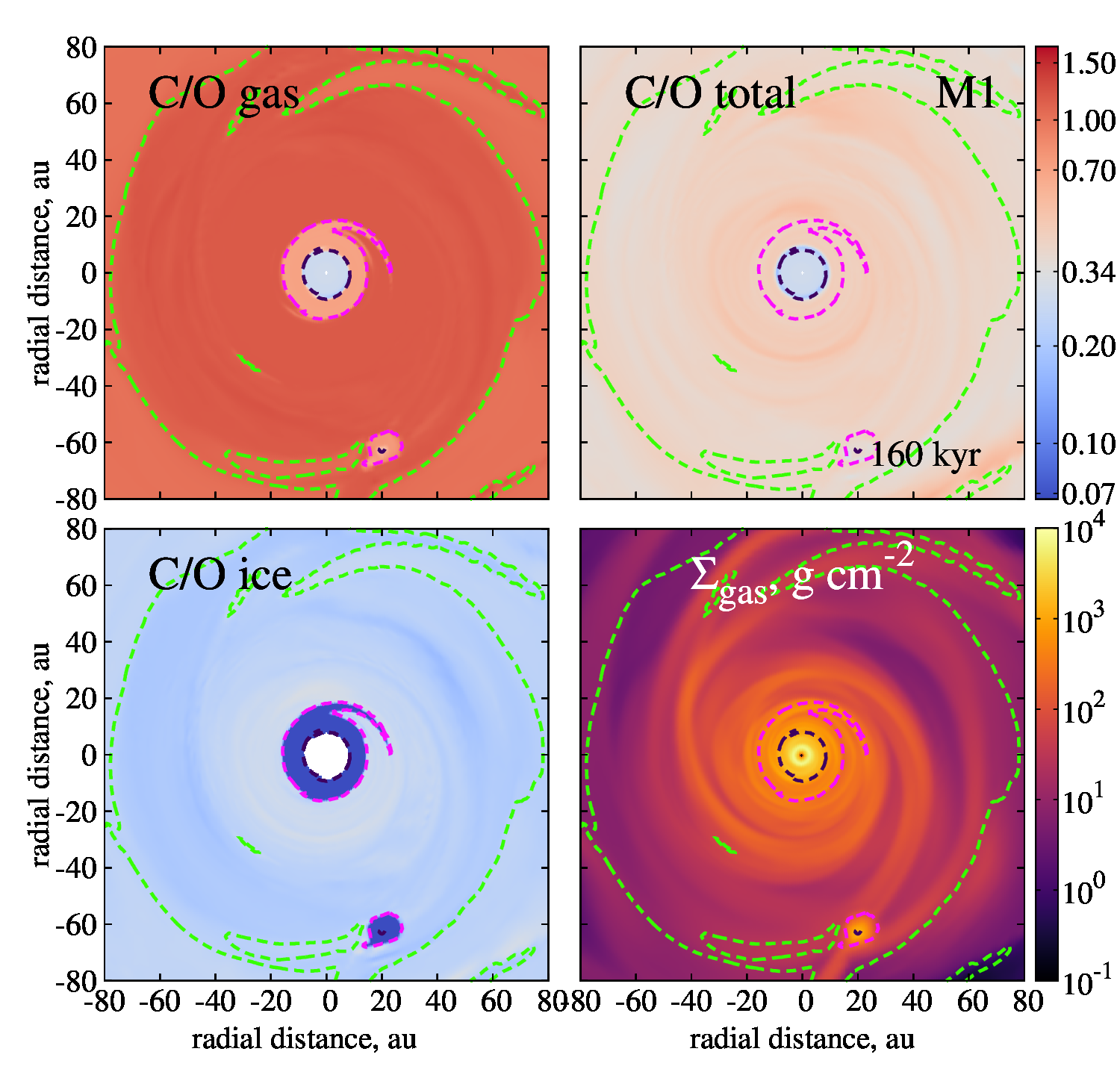}
\includegraphics[width=0.49\columnwidth]{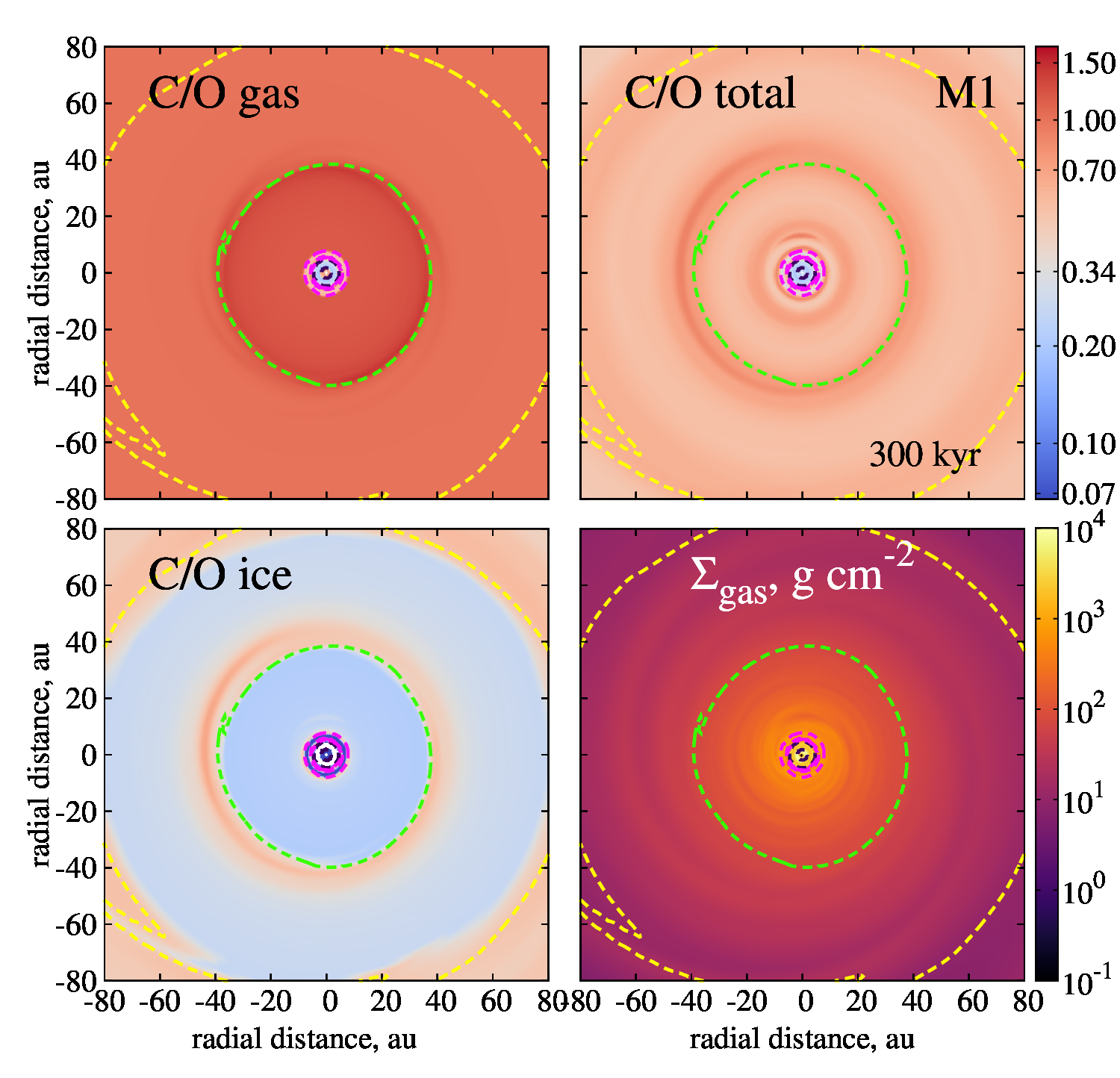}
\includegraphics[width=0.49\columnwidth]{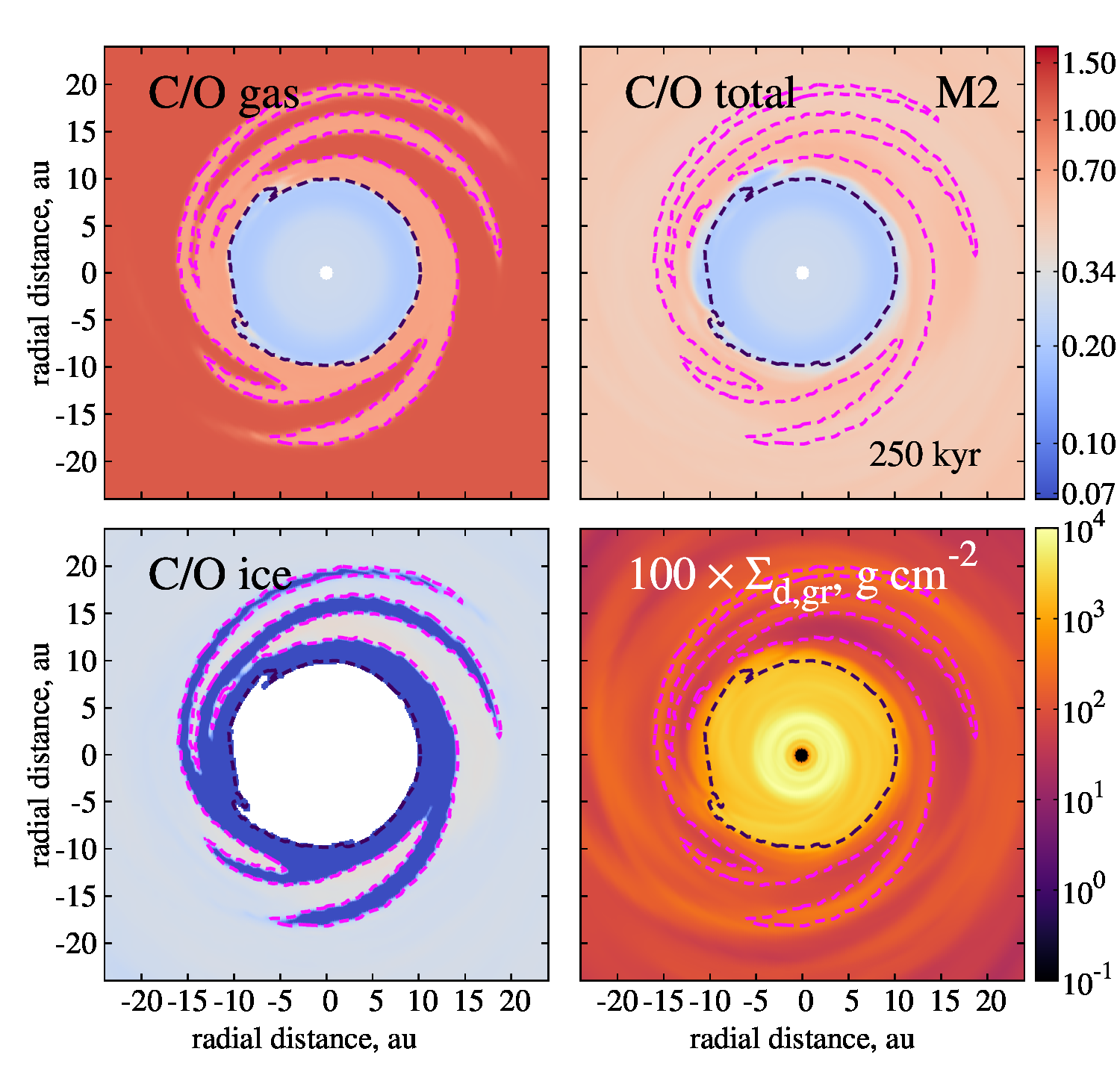}
\includegraphics[width=0.49\columnwidth]{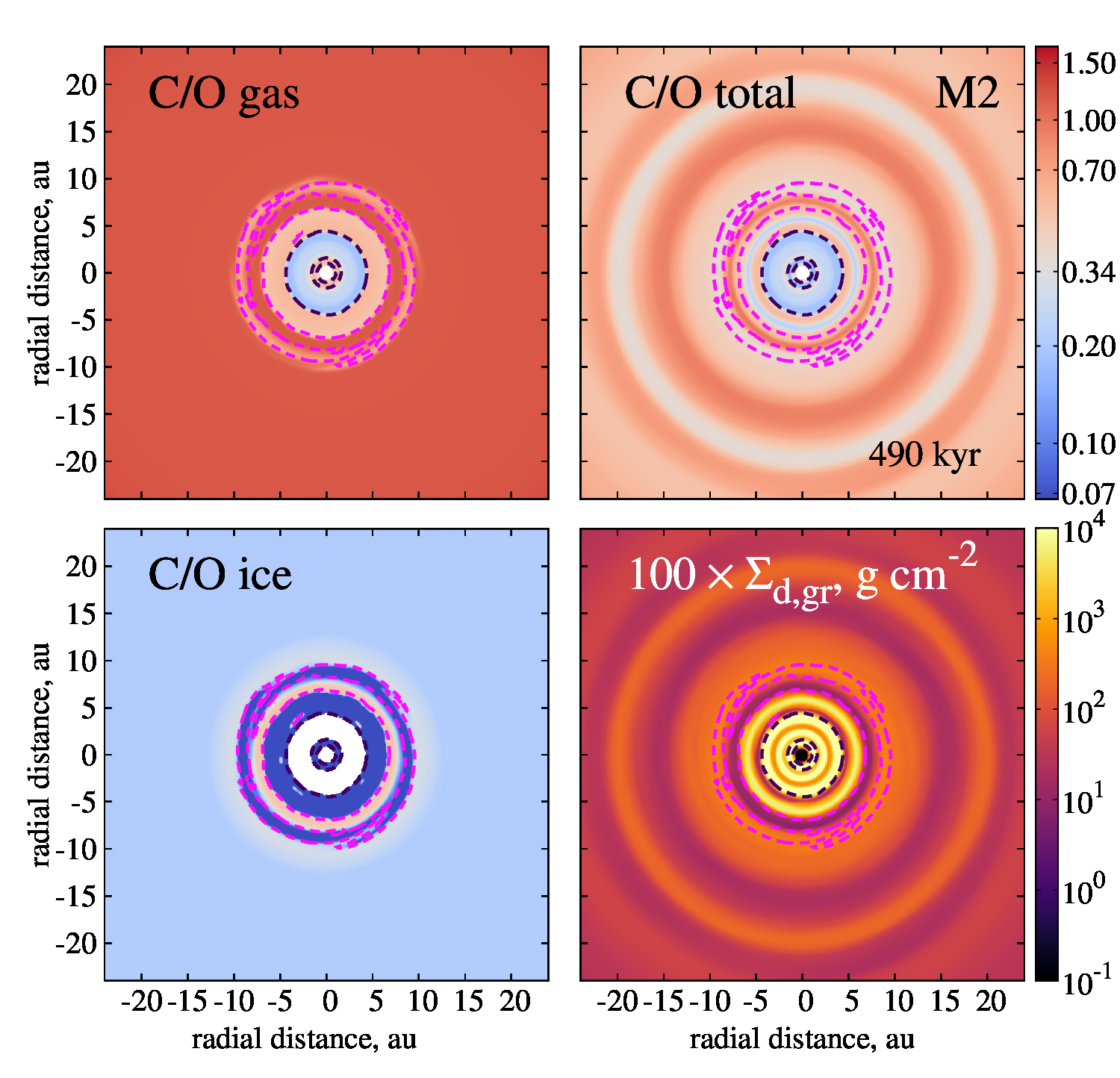}
  \caption{Distributions of C/O ratios and gas/dust surface densities. Model~M1  160\,kyr (upper left) and 300\,kyr (upper right); model~M2 250\,kyr (lower left) and 490\,kyr (lower right). Dotted lines mark the positions of the snowlines for H$_2$O~(dark purple), CO$_2$~(magenta), CH$_4$~(green), and CO~(yellow).}
 \label{fig:c2o_maps}
\end{figure*}

\begin{figure}
\includegraphics[width=\columnwidth]{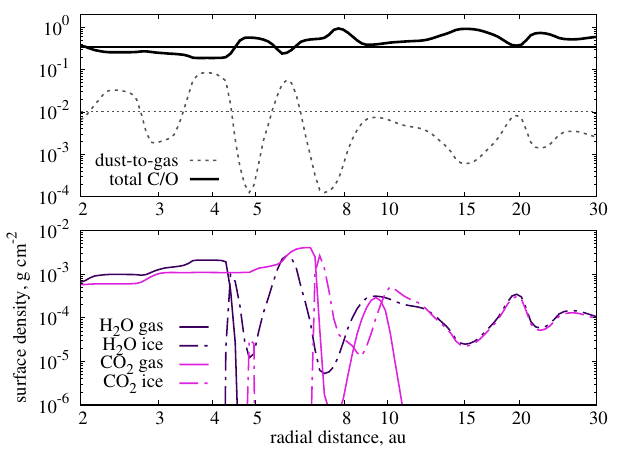}
 \caption{Averaged radial profiles of dust-to-gas ratio, the total C/O ratio, and CO$_2$ and H$_2$O in the gas and in the ice in model~M2 at 490\,kyr. The horizontal lines in the upper panel show the reference values for the C/O ratio (0.34) and dust-to-gas ratio (0.01).}
 \label{fig:M2_radial_snapshot}
\end{figure}

\begin{figure}
\includegraphics[width=\columnwidth]{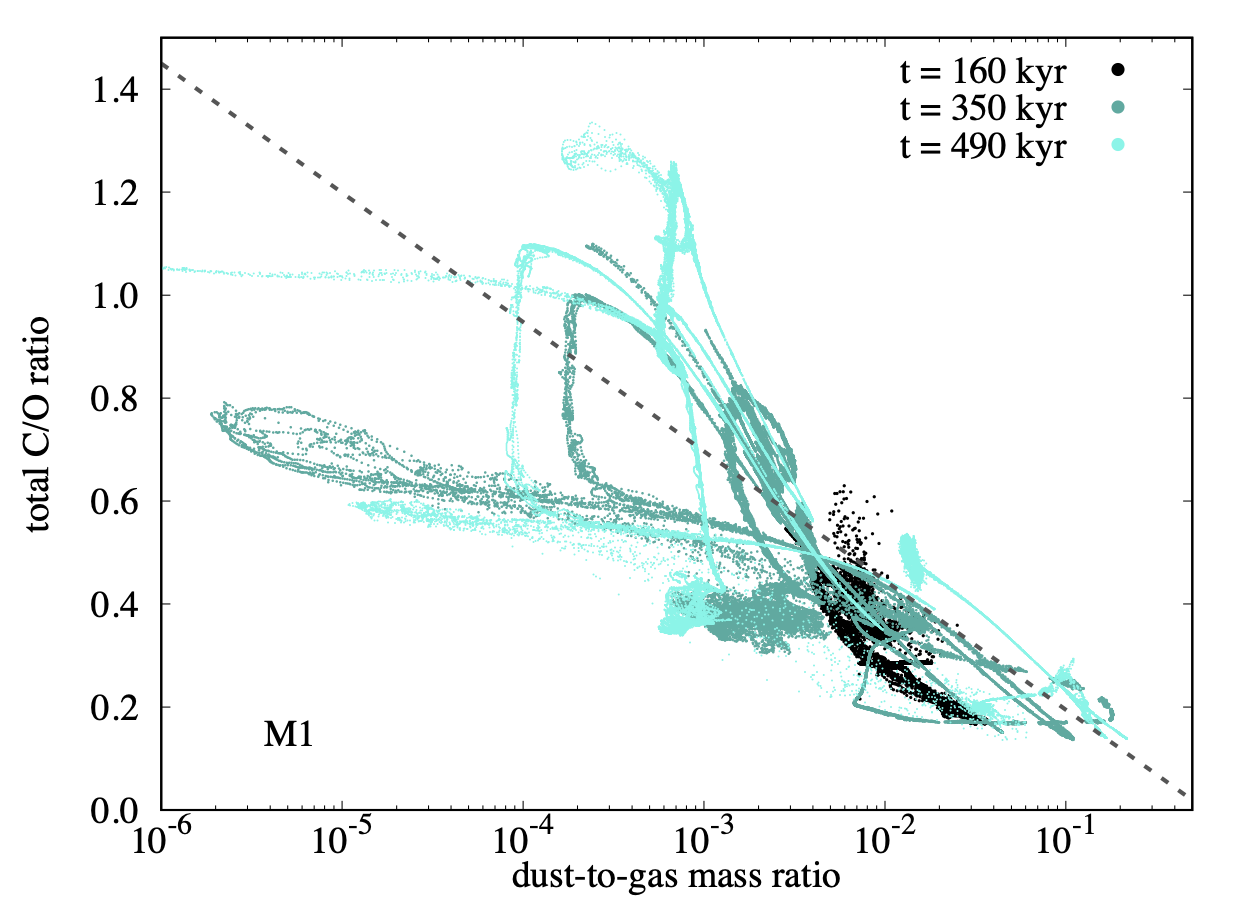}
\includegraphics[width=\columnwidth]{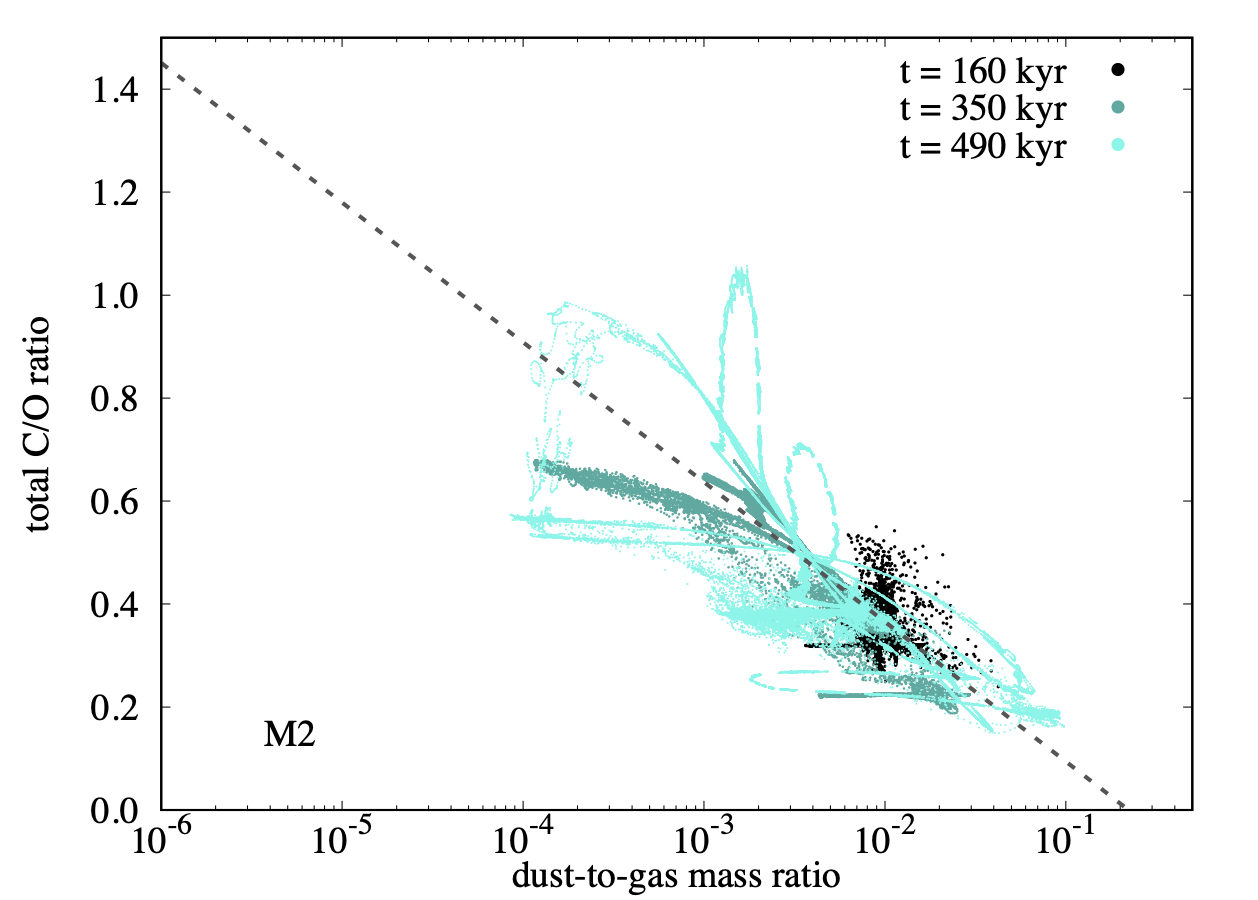}
 \caption{Dependence between total C/O ratio and dust-to-gas mass ratio in models M1 (upper panel) and M2 (lower panel). Three time instances are shown. The dashed line shows the fitted log-linear dependence for $490$\,kyr.}
 \label{fig:anticorr}
\end{figure}

The disc is not axisymmetric even at later stages of its evolution (see Section~\ref{sec:dustgasstructures} and Figure~\ref{fig:maps1}), which is also reflected in the distributions of volatiles and C/O ratios. Examples of 2D distributions of C/O in model~M1 at two time instances are shown in upper panels of Figure~\ref{fig:c2o_maps}. The left-hand side group of panels shows the structure that is characteristic of the earlier phases, the same time instance as the upper row in Figure~\ref{fig:maps1}. The prominent spiral structure and a clump both have their reflection in the C/O ratios. The snowlines of CO$_2$ and CH$_4$ have clearly non-regular shape affected by the spiral pattern of the gas, with blobs of frozen methane inside the main snowline. Inside the clump, the temperature is higher (see Figure~\ref{fig:maps1}), and both CO$_2$ and water are in the gas. The separation between gas-phase and ice-phase C/O ratios is clear, but the total C/O ratio in the clump is similar to the surroundings and is only slightly above the primordial value. The gas-phase C/O ratio inside the clump is around $0.7$, lower than in the surrounding gas at the same radial distances. Similar decrease of C/O ratio indicated by lowered CS/CO ratio was observed in the disc around DR~Tau~\citep{2024arXiv240701679H}. Lifetime of the clump (several orbital periods) is too short for significant differentiation between gas- and ice-phase composition to develop, mainly because of insufficient numerical resolution. Focused studies with higher resolution are needed to explore the C/O ratio in the clumps as precursor of giant planets formed via disc fragmentation.

The upper right-hand side group of panels in Figure~\ref{fig:c2o_maps} features a later stage of the disc evolution in model~M1. By 300\,kyr, the accretion rate and the average positions of the snowlines are stabilised (see Figure~\ref{fig:c2o_m1_m2}). The spiral structure in the gas is much weaker but still present at $r>10$\,au. The CO snowline is established at $\approx 80$\,au. The 2D shape of the snowlines is more circular at 300\,kyr, particularly for the less volatile CO$_2$ and H$_2$O. The effect of spirals is still evident in the contours of the CO and CH$_4$ snowlines. At the outer side of these snowlines, at approximately 40 and 80\,au, respectively, the C/O in the gas has local maxima, which are also seen in Figure~\ref{fig:c2o_radial}. The one at 40\,au also has a clearly spiral shape, repeating the pattern of the gas distribution. The radial span of these peaks is of the same scale as the dispersion of the respective snowline distance from the star. For example, at the CH$_4$ snowline the C/O peak is $\approx10$\,au wide, and the snowline is at 28~to 37\,au distances. Similar accumulation powered by diffusion of water vapour was shown, e.g., by \citet{2017A&A...608A..92D}. In addition to diffusion, in our modelling, the species are delivered outward from the snowline due to the dynamic shape of the snowline and two-dimensional movement of gas and dust \citep[see][and Molyarova et al., in prep]{2021ApJ...910..153M}.

In more massive model~M2, the shape of the snowlines remains complex for longer times. Examples of C/O distributions in~M2 are shown in the lower panels of Figure~\ref{fig:c2o_maps}. The radial scale is chosen so that CO$_2$ and H$_2$O snowlines are seen in more detail. At 250\,kyr, even the grown dust distribution still has spiral pattern, and CO$_2$ snowline is following its complex multi-armed shape. The accumulation of CO$_2$ at the snowline is also very efficient, but it does not strongly affect the C/O ratios, as it drives them to the value 0.5 of the CO$_2$ molecule, which is close to the intrinsic value. The complex-shaped region between these snowlines has the ice-phase C/O~$\approx 0.7$. This pattern moves and changes its shape, thus affecting all radial distances between 10~and 20\,au. The total C/O ratio is slightly above the ambient value, approaching 0.5, because CO$_2$ in both phases begins to accumulate in this region.

At later times, the C/O distribution becomes more complex due to the presence of disc substructures, particularly the dust rings. The lower right-hand side group of panels in Figure~\ref{fig:c2o_maps} show this in more details. First, the temperature and density variations across the dust rings lead to the formation of multiple CO$_2$ and H$_2$O snowlines. An additional annulus of icy CO$_2$ appears in a relatively cold region between the dense dust rings at $6-7$ and $8-13$\,au. The dust rings are warmer due to higher optical depth and active heating by disk internal sources, and the inner edge of the 6\,au ring is warm enough to sustain gas-phase CO$_2$. Second, the accumulation of icy dust grains in the rings alters the total C/O ratio. The total C/O ratio anticorrelates with the distribution of grown dust grains: inside the dense rings it is close to the initial value of 0.34, while between the rings, it is higher and reaches $0.8-0.9$. At the same time, neither ice-phase, nor gas-phase C/O ratio displays any noticeable variations at 12-25\,au, but they do have variations at $r< 12$\,au, following the dust ring pattern.

The total C/O ratio is defined by the combination of the ice- and the gas-phase component. Their relative contribution is proportional to the dust-to-gas mass ratio. This is illustrated in Figure~\ref{fig:M2_radial_snapshot}. 
Within the rings, the total C/O is dominated by ices of oxygen-rich species CO$_2$ and H$_2$O, particularly on grown dust accumulated in the pressure maxima. For example, there is a lot of water ice at  $6-7$\,au, and its contribution to the total C/O is weighted with high dust-to-gas ratio of almost $0.1$, so the resulting total C/O is lower than the initial value. At the same time, at $\approx5$\,au, dust-to-gas ratio is around $10^{-4}$. The dominant species there are in the gas, ice surface densities are $1-2$ orders of magnitude lower, so the total C/O ratio goes up.
In the wide dust ring at $9-13$\,au, both H$_2$O and CO$_2$  contribute, although at the warmer inner edge of the dust ring CO$_2$ is sublimated. The value of total C/O is approaches 0.34, also elevated by the presence of gas-phase CO and CH$_4$. Beyond $\approx10$\,au, both H$_2$O and CO$_2$ are frozen, and the variations of the total C/O ratio are clearly anticorrelated with the dust-to-gas ratio and the position of the rings. Between the dust rings, where contribution of ices is low, the C/O is determined by CO and CH$_4$ gases, and reaches values of $\approx1$.

The anticorrelation between the total C/O ratio and dust-to-gas mass ratio is an interesting finding. It is illustrated in Figure~\ref{fig:anticorr} for both models. Only the points within the disc are shown, with $\Sigma_{\rm gas}>0.1$\,g~cm$^{-2}$.The pattern obviously changes with time, but the anticorrelation persists. Model~M1 demonstrates wider variety of C/O and dust-to-gas ratios. The disc points are grouped in tangled curved "branches", some of them steeper than others. The ``width'' (or spread) of these branches is determined by the azimuthal substructures. In the axially symmetric parts of the disc the values of dust-to-gas ratio and the total C/O are similar at a given radial distance. Different branches, which can be closer to vertical or horizontal orientation, are the result of the radial variations in the ice-phase C/O ratio and in ice fraction relative to the dust silicate cores. Horizontal branches correspond to weak or absent anticorrelation. For example, near the snowlines of carbon-rich species, C/O in the ice is high and close to  that of the gas, which decreases the effect of dust mass fraction on the total C/O. In areas with no ices, the anticorrelation is also irrelevant, because the total C/O is determined entirely by the gas phase. Vertical branches, on the contrary, correspond to the strongest anticorrelation effect, which is expected in the regions between the snowlines, where ice-phase C/O is the lowest. The identified anticorrelation in Figure~\ref{fig:anticorr} is similar to the results of chemical population synthesis modelling \citep{2019A&A...632A..63C,2020A&A...642A.229C} showing that the more solids a planet accreted in the disc, the lower the C/O ratio is in its atmosphere.

We fit the data for $t=490$\,kyr with a linear law (taking the logarithm of dust-to-gas ratio) and obtain the following fits: $\mathrm{C/O} = -0.056 -0.25\log_{10} \left(\xi\right)$ for model~M1, and $\mathrm{C/O} = -0.18 -0.27\log_{10} \left(\xi\right)$ for model~M2. Here, $\xi$ is the dust-to-gas mass ratio. The correlation coefficients are $-0.54$ for M1 and $-0.57$ for M2. At the shown earlier times (160 and 350\,kyr), the correlation coefficient changes between approximately $-0.9$ and $-0.5$, which indicates noticeable anticorrelation throughout the disc evolution.

We note that these C/O ratios only include the volatile component, without the contribution from the refractory material. Although the refractory material is typically considered as silicates, which are rich oxygen, it contains a significant amount of carbon, with the resulting $\mathrm{C/O}\approx0.5$~\citep[see Table~2 in][]{2021ApJ...906...73H}. This solid carbon can be subject to carbon grain destruction \citep[][]{2010ApJ...710L..21L,2017A&A...606A..16G,2019ApJ...870..129W}, but this process should be treated separately, as it also affects the gas-phase carbon abundance. Taking refractory cores into account should affect the dependence between total C/O ratio and dust-to-gas ratio, as adding more rock would make the C/O ratio closer to $0.5$. Thus the degree of the anticorrelation must be affected by the composition of rocky cores, but the anticorrelation itself should remain even when refractories are included, because the $\mathrm{C/O}\approx0.5$ is still lower than typical $\mathrm{C/O}$ of the gas in most of the disc ($\gtrapprox1$).

Dust rings are detected in the majority of the observed protoplanetary discs \citep{2018ApJ...869...17L,2018ApJ...869L..42H}. They are considered as a plausible sites of planet formation \citep{2015A&A...579A..43C,2017A&A...606A..80Y,2021ApJ...919..107L,2022ApJ...937...95L,2023MNRAS.518.3877J}. The anticorrelation between the total C/O ratio of the volatiles and dust-to-gas mass ratio that we point out is a logical consequence of ices having typically lower C/O ratios and being attached to dust grains. If planets are formed in the dust rings with high dust-to-gas ratios ($>10^{-2}$), either exclusively from solids, or with the inclusion of the dust component, this would imply that their material initially has lower C/O ratio of $\approx0.5$ and below. To reach higher C/O ratios up to unity and above, which are observed in many exoplanets, these planets would need to migrate and accrete carbon-rich gas from regions other than their immediate formation sites inside the dust rings. In case if planet formation occurs independently of the dust rings, e.g. in the GI, their material is not determined by this anticorrelation.

\section{DISCUSSION}
\label{sec:discussion}

Our simulations present a wide range of C/O ratios in the disc in different phases evolving with time. For the atmospheres of giant exoplanets, a variety of C/O ratios were retrieved, too. Here we can compare them to identify the disc regions and times where the chemical and physical conditions for planet formation are consistent. Most of the exoplanets for which the atmospheric composition was retrieved have super-stellar C/O ratios \citep{2023AJ....166...85H,2024arXiv240509769W}, which draws more attention to carbon-enriched areas. They are suggested to form by core accretion, and accreting mostly the gas, which is typically more carbon rich (beyond water snowline). A lot of planets are observed to have stellar or slightly super-stellar C/O \citep[e.g.,][and many others]{2020A&A...640A.131M,2021Natur.595..370Z,2024AJ....167..110S,2024arXiv240511027S,2024arXiv240412363N}. One way to form such planets is gravitational instability, which includes solids and gas together, thus undifferentiated matter is suitable for producing planets with stellar C/O. Disc fragmentation  to clumps due to GI requires particular conditions \citep{2010MNRAS.406.2279M,2013A&A...552A.129V}, and the direct collapse of gravitationally unstable clumps tends to produce rather massive objects \citep[e.g. $\approx5$\,$M_{\rm J}$ planets and  $\approx60-70$\,$M_{\rm J}$ brown dwarfs, see Figure~4 in][]{2013MNRAS.433.3256V} at larger radial distances \citep[$>10-100$\,au, see][]{2013MNRAS.433.3256V,2016ARA&A..54..271K}.  GI can also assist the assemblage of planetary cores  \citep{2010MNRAS.408L..36N,2010MNRAS.408.2381N,2014MNRAS.440.3797N,2019A&A...631A...1V}. 
A planet formed through core accretion can also accrete planetesimals, which can be covered with ice, and enrich the atmosphere with oxygen, making the C/O ratio close to the initial stellar value. There are particular exoplanets, where lower than stellar C/O ratio is observed in the atmosphere, such as $\beta$~Pic~b \citep{2020A&A...633A.110G,2024ApJ...964..168W}, HD~209458~b \citep{2024ApJ...963L...5X}, or HD~189733b    \citep{2024arXiv240706163F}. Such planets need even more enrichment in ices with low C/O, which makes the regions with low C/O in the ice also more attractive sites for planet formation.

\begin{figure}
 \includegraphics[width=\columnwidth]{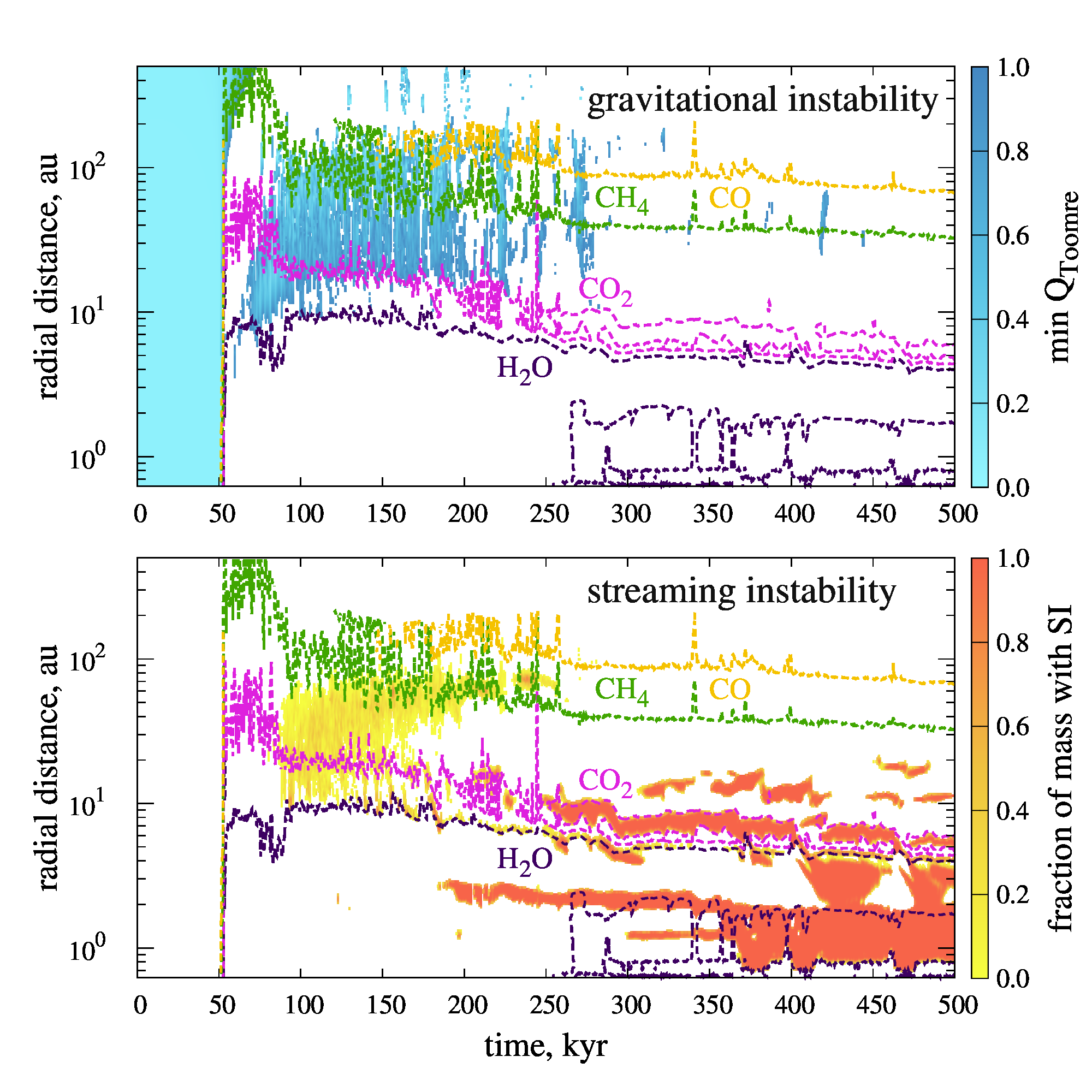}
 \caption{The disc regions where the conditions for GI and SI are fulfilled in model~M1. The regions and times where there is no instability are shaded in white. In the upper panel, the colour indicates the minimum value of $Q_{\rm Toomre}$ at a given radius, if  $Q_{\rm Toomre}\leq1$. In the lower panel, the colour indicates the fraction of mass at a given radius where SI can be triggered according to \citet{2021ApJ...919..107L} criterion. Positions of the snowlines are shown for reference in dashed lines.}
 \label{fig:SI_GI_regions}
\end{figure}

Gravitational instability implies that the planet forms from a mix of gas and dust \citep{1974Icar...23..319B}, this is why it is suitable to explain the formation of planets with solar, or unaltered C/O ratios. In our modelling, GI would be associated with the total C/O ratio, which we find to be significantly variable, too. For GI to form a planet with a primordial C/O ratio, it has to occur during the first 100\,kyr after the disc formation. At later times, the total C/O ratio changes, and the only region with the primordial C/O ratio is the very outer disc parts, at $>100$\,au, which is the part of a protoplanetary disc, where conditions for GI are the most consistent with the observed properties of these objects \citep{2005ApJ...621L..69R}. Planet formation through GI is indeed more likely at earlier evolutionary stages, when gas surface density is higher \citep{2010apf..book.....A}. We can highlight the areas where planet formation via GI is possible in our modelling as the regions where $Q_{\rm Toomre}\leq1$. They are shown in the upper panel of Figure~\ref{fig:SI_GI_regions} for model~M1. These regions appear between $\approx10-100$\,au before $\approx300$\,kyr. However, at later times, clumps could appear in the disc as a result of an external perturbation, such as a stellar flyby \citep{2010ApJ...717..577T}. In this case, the planet would be formed from the material with altered C/O ratio, most probably with elevated amount of carbon, as the regions outside $5-10$\,au are typically more gravitationally unstable. This means that GI can produce planets with super-solar C/O ratios, if it is induced by external influence at later stages of disc evolution.

Core accretion is another most widely discussed scenario of giant planet formation. Accretion of gas should produce atmospheres with the C/O ratio close to the one in the gas phase of protoplanetary disc. However, dust grains are also accreted, so pebble and planetesimal accretion can enrich the atmosphere in volatile components \citep{2016ApJ...832...41M,2023A&A...679L...7D}. This makes atmospheric C/O ratio closer to the ice-phase C/O, but in case of gas giants, the amount of the solids needed to compensate the prevalence of carbon in the gas should be quite high, up to hundreds of Earth masses \citep{2020A&A...633A.110G}. The C/O ratios in exoplanetary atmospheres are often interpreted in terms of pebble accretion, so planets with stellar C/O ratios are assumed to form in the environments where solid phase C/O is unprocessed and thus close to the initial value. One of such locations is beyond CO snowline, where most of the carbon- and oxygen- bearing material is in the ice, e.g. for Jupiter~\citep[e.g.][]{2019AJ....158..194O,2021A&A...651L...2O}. In our models, this is rather the vicinity of the  CO$_2$ snowlines, and the pebbles beyond the CO snowline are mostly covered with carbon-rich CO ice \citep{2024MNRAS.530.2731T}. Additionally, the CO snowline is typically very far from the star ($>40$\,au), so such scenarios must rely on planet migration to obtain their current location. Interpretations relying on chemical modelling extend this region to include the area beyond CO$_2$ snowline due to additional chemical processing of CO in this region, e.g. HR~8799e~\citep{2020A&A...640A.131M}. This puts milder constraints on the original distance from the star where pebbles should be accreted and requires less migration. Modelling of planet formation and migration including pebble and gas accretion puts the formation location of planets with super-solar C/O ratios beyond water and CO$_2$ snowlines \citep{2022A&A...665A.138B}.

In our modelling, the C/O in the ice is close to initial value in the regions beyond CO snowline, excluding the area of CO accumulation. Between CO and CO$_2$ snowlines, it is lower, as our model does not include chemical processes apart from adsorption and desorption. However, there is another region with C/O in the ice close to initial. The vicinity of primary water snowline and the ring induced outside of it has values of C/O in the ice only slightly above the initial value of 0.34. It is surrounded by the snowlines of CO$_2$. This region could be another favourable location for forming planets with the stellar C/O. As it is situated closer to the star, it would imply less migration.

Rare planets with lower than stellar C/O ratio, such as $\beta$~Pic~b \citep{2020A&A...633A.110G,2024AJ....167...45R}, HD~209458~b \citep{2024ApJ...963L...5X}, HD~189733b \citep{2024arXiv240706163F}, or KELT-1~b, Kepler-13A~b and WASP-79~b \citep[less precisely determined, see][]{2023AJ....166...85H}, need to have accreted a lot of oxygen-rich ice. Therefore, they are more likely to accrete solid material in the regions with the lowest ice-phase C/O ratios. The most suitable region would be at the distances between H$_2$O and CO$_2$ snowlines, where ice mantles are made of pure water. However, in our modelling results, this region is very small, typically only a few au wide, as the snowlines are close to each other. This is because of steep temperature profile in this region, which is a result of the significant contribution of non-irradiation heating mechanisms, particularly viscous heating. Beyond CO$_2$ snowline, there are also regions with relatively low ($0.2-0.3)$ C/O in the ice, but much more solids need to be accreted in such areas to compensate for the excess of carbon from the gas.

Let us summarise the above constraints on planet formation locations and mechanisms implied by our simulated C/O ratios. Core accretion is suitable for forming planets with high C/O ($\approx 1$) in the atmosphere around the snowlines of CO, CH$_4$ and CO$_2$, or anywhere beyond CO$_2$ snowline if they did not accrete much solids. Planets with stellar or slightly super-stellar C/O ratio need to accrete (a lot of) oxygen-rich solids to compensate their initially high C/O inherited from the gas. The locations where this is possible is between CO$_2$ and CH$_4$ and between CO and CH$_4$ snowlines. Planets with low C/O ratio could accrete ices between H$_2$O and CO$_2$ snowlines. Snowlines are favourable planetesimal formation sites, so the planets that accreted planetesimals/pebbles there can have altered C/O ratios. The values will be lower than the initial if they form at the water snowline, and higher if they form at the snowlines of carbon-rich species. At the same time, to obtain planets with stellar C/O formed at the snowlines, these planets would need to migrate and accrete matter in different regions of the disc to make their C/O ratio close to the initial stellar value. Alternatively, planets with stellar C/O ratio can form via disc fragmentation through GI at earlier stages. Dedicated modelling of planet formation accounting for evolution of dust and volatiles is necessary to put more particular constraints on planet formation scenarios.

\begin{figure}
\includegraphics[width=\columnwidth]{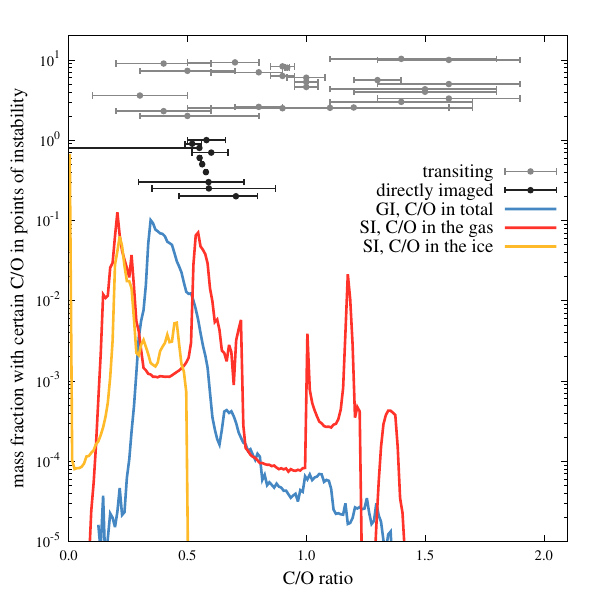}
 \caption{Distribution of C/O ratios in the regions where gravitational and streaming instabilities are triggered. For GI, total C/O ratio is shown, for SI, the C/O ratios in the ice and in the gas. Black and grey points show the observed C/O ratios in two populations of exoplanets, the data is adopted from \citet{2023AJ....166...85H}.}
 \label{fig:C2O_planet_formation}
\end{figure}

Planetesimals play an important role in delivering the ice-phase elements to planetary atmospheres. 
To form planetesimals, additional physical process is needed, such as streaming instability \citep[SI, see][]{2005ApJ...620..459Y}, which is not explicitly included in our modelling because of insufficient numerical resolution and simplified vertical disc structure. However, we can post-process the simulation results to check if the conditions for SI are fulfilled in some regions of the disc where dust-to-gas ratio and dust size are enhanced, following \citet{2024arXiv240416151V}. Dense dust rings forming at later stages (see Section~\ref{sec:dustgasstructures}) seem to be an ideal location for triggering the SI, which would ultimately lead to formation of planetesimals and then planets in the disc. Triggering the SI requires specific relations between local dust-to-gas ratio and Stokes number  \citep{2017A&A...606A..80Y}. The criteria vary depending on the model, we adopt them from \citet{2021ApJ...919..107L}. Another criteria would be the requirement of volume density of dust to exceed that of gas in the midplane \citep{2005ApJ...620..459Y}. We do not apply it, as in our modelling, the residual value of $\alpha$ is $10^{-3}$ which makes this condition unreachable outside of the dead zone. The regions in model~M1 where the conditions  of \citet{2021ApJ...919..107L} are satisfied are shown in the lower panel of Figure~\ref{fig:SI_GI_regions}. Most of the suitable regions are in the inner disc ($r<20$\,au) inside the dust rings, and appear after 200\,kyr. However, there are some suitable regions between $10-100$\,au at earlier times, where SI could be triggered in the spirals.

The regions where GI and SI are possible shown in Figure~\ref{fig:SI_GI_regions} are separated in space and time, and they have different characteristic C/O ratios. We can sum up all the volatiles in these regions (throughout the disc lifetime) to assess typical C/O ratios of the planet-forming material. For GI, we exclude the pre-disc phase ($t<53$\,kyr) and consider the total C/O ratio, assuming both gas and solids are included in the forming planet. For SI, we separate the gas- and ice-phase C/O ratios. The formed planetesimals would only include the ices, however, if they form the planetary cores, these cores would also accrete gas. We note that the composition of the rocks, which are typically carbon-rich, is not included in our assessment. The resulting distributions of the C/O ratios in planet-forming regions are shown in Figure~\ref{fig:C2O_planet_formation}. For GI regions, the C/O distribution has a distinct and relatively narrow peak around 0.5. It is slightly higher than the initial value of $0.34$. For SI regions, the ice-phase C/O is below 0.5, with major peaks at 0 and $\approx0.2$, and the gas-phase C/O has a broad distribution with multiple peaks between $\approx0.2-1.4$. Distributions of C/O ratios in the regions where GI and SI can be triggered are noticeably different.

It was shown by \citet{2023AJ....166...85H} that there are two different populations of C/O ratios observed in giant exoplanets. They find that directly imaged exoplanets have $\mathrm{C/O}\approx0.5-0.8$, while transiting hot Jupiters have a wider variety of C/O ratios \citep[$\approx0.3-1.7$, see Figures~12 and~13 in][]{2023AJ....166...85H}, and suggest that these two populations could have different formation pathways. We add the C/O data of exoplanetary atmospheres compiled in Table~3 of~\citet{2023AJ....166...85H} to Figure~\ref{fig:C2O_planet_formation} (with arbitrary position at the $y$-axis). The narrow distribution of C/O ratios in the regions with GI is in step with the distribution of directly imaged exoplanets, albeit with a slightly shifted value due to our assumed initial conditions, while the wide range of C/O values in the regions of SI matches the variety of C/O ratios in transiting exoplanets. This could suggest that directly imaged exoplanets could form as a result of gravitational instability, which is also in line with their typically higher masses and orbital separations. At the same time, the transiting hot Jupiters could have experienced a lot of migration \citep{1996Natur.380..606L,2018ARA&A..56..175D}, during which they accrete material with a variety of C/O ratios both from the gas and solid phase. It suggests that they could also form in the core accretion scenario. The origin of wide separation planets was also investigated by \citet{2024arXiv240612037B}, based on the comparison with the observed $\mathrm{C/O}>1$ in protoplanetary discs (including the full composition of solids). They conclude that both core accretion and gravitational instability can work as the formation mechanism of these planets.

Apart from (exo)planets, the C/O ratios can be measured for the comets, which present the best preserved sample of the primordial composition of the ices in the Solar System. Spectroscopic measurements of molecular composition in the comae suggest that the C/O ratio of cometary ice is quite low, typically below $0.1$ due to the dominance of water ice \citep{2012ApJ...758...29A,2022PSJ.....3..150S,2022PSJ.....3..247H}. Although most comets are carbon-depleted, there are individual measurements of C/O in comets above 0.5, for example in C/2006~W3~Christensen and 29P/Schwassmann--Wachmann \citep{2012ApJ...752...15O,2022PSJ.....3..150S}, or even close to 1 in C/2016~R2~(PanSTARRS) \citep{2018AJ....156...34W,2019AJ....158..128M}. Additionally, high value of $\mathrm{C/O}\approx 1$ was observed in the interstellar object 2I/Borisov \citep{2020NatAs...4..867B}. Our modelling results show the ice-phase $\mathrm{C/O}=0$ in the vicinity of the water snowline, as well as low values between the CO$_2$ and CH$_4$ snowlines ($\approx0.2$) and between the CH$_4$ and CO snowlines ($\approx0.3$). These are the locations where comets could originate from. However, in the vicinity of the CO$_2$, CH$_4$ and CO ice lines themselves, the C/O ratio in the ice phase is much higher. The fact that carbon-rich cometary ices are extremely rare in the Solar System may indicate that planetesimals formed on ice lines from carbon-rich volatiles do not persist throughout the evolution of a planetary system. This means that they are likely to be included in larger bodies, which favours the snowline-aided planet formation scenarios \citep{2017A&A...608A..92D,2021A&A...646A..14H}. This is also consistent with the abundance of exoplanets with high C/O \citep{2024arXiv240509769W}, which could be formed around the snowlines of carbon-rich species. In the giant planets of the Solar System, the C/O ratios are not well constrained \citep{2024SSRv..220...44M}. However, the existing data suggest rather super-solar values for all giant planets except for Neptune \citep{2024SSRv..220....8C}; for Jupiter, the C/O ratio is assessed as $\approx0.9$~\citep{2004Icar..171..153W,2024Icar..41416028L}.

Our model only considers four most abundant chemical species. However, there can be other more complex molecules in protoplanetary discs, which could affect the balance of carbon and oxygen. The most obvious candidate is methanol CH$_3$OH, which has the abundance similar to methane in the protostellar cores \citep{2011ApJ...740..109O}. It was also observed in a protoplanetary disc around an erupting star V883~Ori \citep{2019NatAs...3..314L}. We do not consider it in the model as its binding energy is close to that of water, thus the snowlines would have similar positions, but the abundance is an order of magnitude lower. However, it could somewhat increase the local C/O ratio in the inner regions where there are no other carbon-bearing species, such as the ice in the  ring at 1\,au. Including methanol would alter the distribution of the C/O ratio. Interactions between the ices considered in the model could also affect the results. As was recently shown by \citet{2024arXiv240616029L}, trapping of volatile species inside the mantles of less volatile ices could have a significant impact on the distribution of C/O ratios.

Another important process missing in our modelling is gas-phase and surface chemical reactions. They could significantly affect the distribution of C/O ratio in the gas and in the ice, particularly with high level of cosmic ray ionisation \citep{2016A&A...595A..83E} or if carbon grain destruction is considered \citep{2019A&A...627A.127C}.
One particular mechanism is the transformation of CO to CO$_2$ on the surface of dust grains, which can lead to the depletion of CO from both gas and ice phases between CO and CO$_2$ snowlines \citep{2017ApJ...849..130M,2018A&A...611A..80B}. 
Considering this mechanism can change the conclusions about planet formation location \citep{2020A&A...640A.131M}. Nevertheless, radial variations of the C/O ratio are necessary to explain molecular emission of  discs with gaps \citep{2024arXiv240510361L}, and they can only be result of dust dynamics.
In order to more consistently describe the distribution of molecules and elements in the disc, the models combining dust evolution and dynamics with more complex chemistry treatment are necessary.

Our simulations adopt the thin-disc approximation and focus on the midplane of the protoplanetary discs, in order to capture the essential physics of self-gravity, thermal balance, and dust evolution in a global modelling within reasonable computational times. This means that some relevant processes connected with the vertical structure are inevitably excluded. For example, vertical mixing and dust settling affect the C/O ratio in the upper layers of the disc \citep{2018ApJ...864...78K,2020ApJ...899..134K}. Dust settling is implicitly included in our modelling through separate scale heights of drown dust and gas (as well as small dust), affecting dust number density in the midplane. However, this approach does not allow to reproduce vertical stratification in dust properties and chemical composition, which is particularly relevant for the interpretation of molecular observations. Vertical structure is also relevant for the accretion of matter on forming giant planets, which should proceed in 3D manner through meridional flows \citep{2014Icar..232..266M}. \citet{2020A&A...635A..68C} showed that the C/O ratio in the atmospheres of giant planets is rather affected by the composition of the molecular layer than that of the midplane.

\section{Conclusions}
\label{sec:conclusions}

In this work, we studied the distribution of volatiles in a viscous self-gravitating protoplanetary disc with dust evolution using a thin-disc hydrodynamic code FEOSAD~\citep{2018A&A...614A..98V,2021ApJ...910..153M}. We calculated the C/O elemental ratio in the gas, in the ice, and in total, identified the key properties of the distribution of elements over 500\,kyr of disc evolution and considered their implications for planet formation theory. Our main findings can be summarised as follows.

\begin{itemize}

    \item The simulated C/O ratios in the regions where GI and SI conditions are fulfilled are consistent with the C/O ratios in two populations of exoplanets possibly formed in different mechanisms pointed out by \citet{2023AJ....166...85H}. We show that narrow C/O distribution of directly imaged planets is consistent with their formation via gravitational instability, while a variety of C/O in transiting hot Jupiters is in line with their migration through varying C/O conditions after the formation via either core accretion or GI.
    
    \item The lower C/O ratio in the ice between the CO$_2$, CH$_4$ and CO snowlines is consistent with the typical composition of Solar System comets, while the higher value of $\mathrm{C/O}\approx 0.5-1$ at these snowlines corresponds to the composition of rare carbon-rich comets. This may indicate that matter from the snowlines is hardly preserved during the evolution of the disc and planetary system, possibly due to the inclusion in to planets.

    \item The distribution of volatiles is affected by the disc substructures, such as rings and spirals, as well as by dust radial drift. Variations of physical conditions create multiple snowlines of CO$_2$ and H$_2$O inside $10$\,au. Dust drift of icy grains brings the volatiles from the outer to the inner disc, enriching the inner disc with both C and O. It also creates a radial gradient of the total C/O ratio: its value is around 0.2 where water is not frozen, and $0.6-0.9$ where it is icy, compared to the initial value of $0.34$.
    
    \item Volatiles accumulate at their snowlines in both ice and gas phases due to the combined effect of dust drift and azimuthal variations of gas and dust radial velocities in a self-gravitating, non-axisymmetric disc. The species with low initial abundances, such as CH$_4$ (or methanol not considered here), can significantly affect C/O ratio, as their accumulation at the snowline creates a bump in C/O in all phases above $1.0$. The total mass of the model affects the timescales and the magnitude of the accumulation by a factor of two.

    \item Forming planets can accrete gas with $\mathrm{C/O}>1$ beyond CO$_2$ snowline, ices with $\mathrm{C/O}\approx0.5-1$ at the CO, CH$_4$ and CO$_2$ snowlines, ices with $\mathrm{C/O}\approx 0.2-0.3$ between these snowlines and ices with $\mathrm{C/O}=0$ between H$_2$O and CO$_2$ snowlines. Planets with stellar C/O would need to migrate through these regions to acquire necessary composition or form via GI at earlier stages from the mixture of gas and dust with unaltered C/O ratio.

    \item Dust-to-gas mass ratio and the total C/O ratio are systematically anticorrelated, because in dust-rich regions the volatile composition is close to that of the ice (which is lower), and in dust-poor regions, gas determines the C/O ratio.
 
 \end{itemize}

The connection between protoplanetary disc components and exoplanets based on their composition should be more thoroughly investigated in  the models focused on the planet formation process. We emphasise that these models should also take into account the effect of dust evolution and dynamics on the distribution of the elements in the planet-forming material. Inclusion of chemical processes and more accurate consideration of the bulk composition of dust grains could also affect the C/O ratios of the planet-forming environment.

\begin{acknowledgement}
We are thankful to the anonymous referee for useful comments that helped to improve the manuscript. The computational results presented have been achieved using the Vienna Scientific Cluster (VSC) and the local computing facility of the Southern Federal University.
\end{acknowledgement}

\paragraph{Funding Statement}
The work is supported by Russian Science Foundation grant 22-72-10029, https://rscf.ru/project/22-72-10029/

\paragraph{Competing Interests}
None

\paragraph{Data Availability Statement}
The data underlying this article will be shared on reasonable request to the corresponding author.

\printendnotes

\printbibliography

\end{document}